\documentclass[]{pasj02} 
\usepackage[switch,mathlines]{lineno} 
\usepackage{url}
\usepackage{arydshln}
\usepackage{xcolor}
\usepackage[normalem]{ulem} %

\newcommand{\phiorb}{\phi_{\text{orb}}}
\newcommand{\Cloudyabs}{\texttt{Cloudy\_abs}}
\newcommand{\pionabs}{\texttt{pion\_abs}}

\Received{}
\Accepted{2026/03/06}

\begin{document} 

\title{XRISM High-Resolution X-ray Spectroscopy of Cygnus X-1\\
--- Orbital and Short-Term Variability of Iron Absorption
}


\author{Kaito Ninoyu\altaffilmark{9,*}\orcid{0009-0003-0640-2828}}
\email{6223525@ed.tus.ac.jp, ninoyu@rikkyo.ac.jp}
\author{Shinya Yamada\altaffilmark{1,*}\orcid{0000-0003-4808-893X}}
\email{syamada@rikkyo.ac.jp}
\author{Natalie Hell\altaffilmark{2,*}\orcid{0000-0003-3057-1536}}
\email{hell1@llnl.gov}

\author{Elisa Costantini\altaffilmark{3}\orcid{0000-0001-8470-749X}}
\author{Oluwashina Adegoke\altaffilmark{4}\orcid{0000-0002-5966-4210}}
\author{Paul Draghis\altaffilmark{5,15}\orcid{0000-0002-2218-2306}}
\author{Ken Ebisawa\altaffilmark{6}\orcid{0000-0002-5352-7178}}
\author{Javier A. Garcia\altaffilmark{4}\orcid{0000-0003-3828-2448}}
\author{Edmund Hodges-Kluck\altaffilmark{7}\orcid{0000-0002-2397-206X}}
\author{Shunji Kitamoto\altaffilmark{1}\orcid{0000-0001-8948-7983}}
\author{Shogo Kobayashi\altaffilmark{8}\orcid{0000-0001-7773-9266}}
\author{Takayoshi Kohmura\altaffilmark{9}\orcid{0000-0003-4403-4512}}
\author{Aya Kubota\altaffilmark{10}\orcid{0000-0002-5413-6304}}
\author{Jon M. Miller\altaffilmark{5}\orcid{0000-0003-2869-7682}}
\author{Misaki Mizumoto\altaffilmark{11}\orcid{0000-0003-2161-0361}}
\author{Tsunefumi Mizuno\altaffilmark{12}\orcid{0000-0001-7263-0296}}
\author{Hiromitsu Takahashi\altaffilmark{13}\orcid{0000-0001-6314-5897}}
\author{Yuusuke Uchida\altaffilmark{9}\orcid{0000-0002-7962-4136}}
\author{Kazutaka Yamaoka\altaffilmark{14}\orcid{0000-0003-3841-0980}}
\author{Sixuan Zhang\altaffilmark{13}\orcid{}}
\author{Ryota Tomaru\altaffilmark{16}\orcid{0000-0002-6797-2539}}
\author{Seoru Ito\altaffilmark{9}\orcid{}}

\altaffiltext{1}{Department of Physics, Rikkyo University, 3-34-1 Nishi Ikebukuro, Toshima-ku, Tokyo 171-8501, Japan}
\altaffiltext{2}{Lawrence Livermore National Laboratory, CA 94550, USA}
\altaffiltext{3}{SRON Netherlands Institute for Space Research, Leiden, The Netherlands} 
\altaffiltext{4}{Cahill Center for Astronomy and Astrophysics, California Institute of Technology, Pasadena, CA 91125, USA}
\altaffiltext{5}{Department of Astronomy, University of Michigan, MI 48109, USA} 
\altaffiltext{6}{Institute of Space and Astronautical Science (ISAS), Japan Aerospace Exploration Agency (JAXA), Kanagawa 252-5210, Japan} 
\altaffiltext{7}{NASA / Goddard Space Flight Center, Greenbelt, MD 20771, USA} 
\altaffiltext{8}{Faculty of Physics, Tokyo University of Science, Tokyo 162-8601, Japan} 
\altaffiltext{9}{Faculty of Science and Technology, Tokyo University of Science, Chiba 278-8510, Japan} 
\altaffiltext{10}{Department of Electronic Information Systems, Shibaura Institute of Technology, Saitama 337-8570, Japan} 
\altaffiltext{11}{Science Research Education Unit, University of Teacher Education Fukuoka, Fukuoka 811-4192, Japan}
\altaffiltext{12}{Hiroshima Astrophysical Science Center, Hiroshima University, Hiroshima 739-8526, Japan} 
\altaffiltext{13}{Department of Physics, Hiroshima University, Hiroshima 739-8526, Japan} 
\altaffiltext{14}{Department of Physics, Nagoya University, Aichi 464-8602, Japan} 
\altaffiltext{15}{MIT Kavli Institute for Astrophysics and Space Research, Massachusetts Institute of Technology, 70 Vassar St, Cambridge, MA 02139, USA}
\altaffiltext{16}{Department of Earth and Space Science, Graduate School of Science, Osaka University, 1-1 Machikaneyama, Toyonaka, Osaka, 560-0043, Japan}




\maketitle

\begin{abstract}
We present the first high-resolution spectroscopy of the black hole high-mass X-ray binary Cygnus~X-1 with XRISM, including orbital-phase-resolved analyses and tentative evidence of short-term variability in the Fe\,K band on second timescales. Using data from the Performance Verification phase in April 2024, we analyzed spectral variability across orbital phases with the Resolve microcalorimeter and the Xtend CCD imager. The unprecedented resolution of Resolve reveals variability in highly ionized Fe absorption lines. The absorption features show orbital-phase-dependent variability in column density, ionization state, and blueshifted velocity, suggesting structural variations in the focused stellar wind along the line of sight. We also find indications of subtle broadening of the neutral Fe emission profile. In addition, intensity-sorted spectroscopy during dip phases suggests possible variability on timescales of a few seconds in the absorption features, consistent with cooler, denser and lower-ionized gas clumps. Although the statistical significance is limited, these results hint that the stellar wind and the X-rays from the accretion disk around the black hole may interact on timescales as short as a few seconds. These XRISM results constrain wind-fed accretion in Cyg~X-1 and highlight Resolve’s capability to probe plasma environments in high-mass X-ray binaries.
\end{abstract}


\section{Introduction}\label{sec:Introduction}
Cygnus X-1 (Cyg X-1) is among the most thoroughly studied black hole binaries; its X-ray history dates back to the pioneering observations by the Uhuru satellite (e.g., \cite{Oda1971, Tananbaum1972}). 
It hosts the most massive black hole known among high-mass X-ray binaries (HMXBs) in the Milky Way, with a mass of $21.2\pm2.2~M_{\odot}$, a companion star mass of $40.6^{+7.7}_{-7.1}~M_{\odot}$, a distance of 2.2~kpc, and a binary inclination of $i = 27^{\circ}.5^{+0.8}_{-0.6}$ \citep{Miller-Jones2021-nt}. 
The companion star, HD 226868, is a blue supergiant of spectral type O9.7~Iab \citep{Walborn1973}. 
The black hole and its companion form a nearly circular binary with an orbital separation of $42.2R_{\odot}$ and an eccentricity of $e = 0.018\pm0.002$ \citep{Orosz2011-ez}, 
orbiting their common center of mass with a period of $\sim5.59~\mathrm{d}$ \cite{Brocksopp1999b}. 
Because of the close separation, continuous mass accretion onto the black hole is driven by the stellar wind from the companion, resulting in an accretion flow. 
Cyg~X-1 is known to display two distinctive spectral states (e.g., \cite{Remillard2006, Done2007, Yamada2013-wg}): the high/soft state and the low/hard state. 
In the high/soft state, the spectrum is dominated by multi-temperature blackbody emission from a cool accretion disk at $kT_{e} \sim 1$--$3~\mathrm{keV}$. 
In the low/hard state, the emission is instead governed by inverse-Compton scattering in a hot corona, with an electron temperature of $kT_{\mathrm{e}} \simeq 100~\mathrm{keV}$ (e.g., \cite{Makishima2008-wn, Tomsick2018-qk}). Its spectrum also includes a reflection component, in which Comptonized photons from the corona are reprocessed by the accretion disk, producing a Compton hump at energies of $\sim20$--$30~\mathrm{keV}$ and the characteristic Fe\,K$\alpha$ fluorescence line (e.g., \cite{Fabian2012, Basak2017}).

Cyg X-1 remains bright and persistent, serving as a key probe of wind-fed mass transfer in high-mass X-ray binaries. Most of the mass accretion occurs through the stellar wind of the O-supergiant companion HD 226868 at a rate of $\sim10^{-6}~M_{\odot}~\mathrm{yr}^{-1}$ \citep{Herrero1995-oh}. The stellar wind is line-driven in the sense of Castor, Abbott, and Klein (CAK; \cite{Castor1975-wm}), where radiation pressure on numerous absorption lines in the stellar atmosphere accelerates the wind, and a fraction of this material is ultimately captured by the black hole. The small distance between the binary stars and the large radius of the companion star suggest that a gas stream is continuously flowing toward the black hole from the inner Lagrange point (L1), as the companion star is close to filling its Roche lobe \citep{Gies1986-bz}. A portion of the stellar wind is focused toward the black hole in a manner similar to Roche lobe overflow, forming a focused wind \citep{Friend1982-qb}. Focused winds in early-type O or B stars are unstable, producing cool and locally dense structures called clumps, as indicated by numerical and hydrodynamic simulations (e.g., \cite{Owocki1988, Oskinova2012, Sundqvist2018}, and references therein).

In Cyg X-1, X-ray dips allow us to investigate the physical structure of the wind. The dips mainly occur around orbital phase $\phi_{\mathrm{orb}} = 0$, i.e., the superior conjunction of the black hole, when soft X-rays are absorbed and the observed flux declines. At this phase, the line of sight crosses the stellar wind, and X-rays emitted near the black hole are partially absorbed as they traverse the wind. Observational studies of the stellar wind in the hard state during the dip phase (including non-dip periods at superior conjunction) have been actively conducted with Chandra observations (e.g., \cite{Miller2005-sw, Hanke2009-rh, Hirsch2019-em, Miskovicova2016-zi}). In its early observations, Chandra successfully resolved Fe absorption features in the low-energy band during a pre-dip phase ($\phi \sim 0.76$) \citep{Miller2005-sw}. \cite{Hanke2009-rh} and \cite{Miskovicova2016-zi} examined non-dip spectra, measuring the orbital-phase dependence of Doppler shifts and column densities in Si and S absorption lines; their results indicate a wind structure with distinct layers. \citet{Miskovicova2016-zi} also reported a P-Cygni profile near $\phi_{\mathrm{orb}}\sim0.5$. Focusing on the dip phase, \citet{Hirsch2019-em} investigated dip-depth variability in the Si and S absorption region (1.6--2.7 keV). Analysis of the ionized Si and S lines revealed that, as the dip deepens, the column density of lower-ionization species increases, whereas that of higher-ionization species decreases. From these results, \citet{Hirsch2019-em} noted the presence of lower-ionization, lower-temperature, or higher-density material in the deeper dip stage. \citet{Harer2023-gc} applied excess-variance spectroscopy to the same observations, reporting complex temporal variability of the dips. Thus, although there has been extensive research on the wind using the band below 3~keV, high-resolution spectroscopy in the band above 6 keV has been performed only in a few cases so far (e.g., Chandra resolved narrow Fe features \citep{Miller2002}). This band corresponds to the Fe region and plays an important role in black hole studies. 

The X-ray Imaging and Spectroscopy Mission (XRISM; \cite{Tashiro2020-iz, Tashiro2025}), launched in 2023, enables high-resolution spectroscopy in the 2--10~keV band and has already yielded innovative observational findings. XRISM is equipped with Xtend \citep{Mori2022-wo}, a soft X-ray imaging spectrometer with a wide field of view and energy range, and Resolve \citep{Ishisaki2022-ci, Sato2023-gr}, which provides the highest energy resolution ever achieved in the 2--8~keV band, at the focal plane of the X-ray Mirror Assembly. The first observation of Cyg~X-1 with XRISM was performed on April 7, 2024 during the Performance Verification phase, and \citet{Yamada2025} (accepted to PASJ, hereafter Y25) reported mainly on the time-averaged spectrum and its interpretation. The spectral analysis with Resolve in that paper successfully investigated the ionized Fe absorption line, revealing an ionization parameter of $\xi \sim 3$, column densities of a few $\times 10^{21}~\mathrm{cm}^{-2}$, and a blueshifted velocity of $\sim100~\mathrm{km~s}^{-1}$. They also reported possible time variability of the absorption features. In the latest XRISM observations, orbital modulation of ionized Fe emission and absorption lines has been reported in Cygnus~X-3 (\cite{XRISM-Collaboration2024-vs}; \cite{Miura2025}) and Centaurus~X-3 \citep{Mochizuki2024-yz}, two other HMXBs similar to Cyg~X-1. 

In this paper, following Y25, we report on the time variability of absorption by highly ionized matter. We present results and discussions on the orbital dependence of the spectrum, utilizing data from both Xtend and Resolve. Black hole spin is a key observable, extensively studied (e.g., \cite{Draghis2024-oz,Zdziarski2024-mi}). Because robust spin estimates require careful continuum modeling, we defer this analysis to a future paper. This paper is organized as follows. In Section \ref{sec:Observation and Data Reduction}, we describe the observations and data reduction. In Section \ref{sec:data analysis and results}, we present the identification of dips and an analysis of the structures of emission and absorption lines. The implications of these results are discussed in Section \ref{sec:Discussion}. Finally, we summarize our findings in Section \ref{sec:Summary}. Unless otherwise noted, the errors quoted in the text and tables, as well as the error bars in the figures, represent the $1\sigma$ confidence level. Note that though \citet{Ramachandran2025-jn} reported lower values for the BH mass and the donor star mass, we adopt the parameters from \citet{Miller-Jones2021-nt}, including LTE models in \citet{Orosz2011-ez}, in this work to maintain consistency with previous observational analyses and model comparisons.

\section{Observation and Data Reduction}\label{sec:Observation and Data Reduction}
The XRISM observation of Cyg~X-1 was conducted from 2024-04-07 16:55:04 to 2024-04-10 13:41:04 (UTC), and the observation ID (OBSID) is 300049010. The observation had an effective exposure of 125~ks. Data reduction was performed using the JAXA pre-pipeline software ``HEAsoft version 6.35,'' the pipeline script ``03.00.013.009,'' and the CALDB8 database, with calibration files ``gen20240315\_xtd20240815\_rsl20240815.'' 

\subsection{Data Reduction of Resolve}
The Resolve observation of Cyg~X-1 was conducted through a beryllium window \citep{Midooka2020-ns}, which limited the observable energy range to above 1.6~keV. Cyg~X-1 is a very bright object; therefore, to mitigate pile-up effects, the observation was made with a neutral density filter. This filter consists of a molybdenum plate that reduces the flux without affecting the spectrum. The Resolve data were screened with the standard pipeline and reduced following the procedure described in the XRISM Quick Start Guide\footnote{\url{https://heasarc.gsfc.nasa.gov/docs/xrism/analysis/quickstart/xrism_quick_start_guide_v2p3_240918a.pdf}}. The source events were extracted from high primary (Hp) grade events and from all pixels except pixel 27, which has calibration uncertainties, and pixel 12, which is designated as a calibration pixel. The Redistribution Matrix Files (RMFs) and Ancillary Response Files (ARFs) were generated for the cleaned source events using \texttt{rslmkrmf} with the ``Large'' option and \texttt{xaarfgen}, respectively. The exposure map required to generate the ARF was created using \texttt{xaexpmap}. We confirmed that the response files (RMF and ARF) were not time-dependent; for this analysis, we used the response files created from the entire dataset, as well as those for time-resolved spectra. For details of the gain calibration in this observation, see Y25. 

\subsection{Data Reduction of Xtend}
The Xtend observation was operated in \texttt{WINDOW2BURST} mode. In this mode, the exposure time is reduced to $1/8$ ($4\times1/8$~s), and only 0.0620352~s is read out (burst mode). This mode is useful for preventing pile-up that occurs when observing bright objects such as Cyg~X-1 with CCDs. The operational bandpass, extending from 0.4 to 13.0~keV, has been empirically confirmed \citep{Noda2025}. The CCD design of Xtend has also resulted in a low background in this energy band.

Because Cyg~X-1 is exceptionally bright, the pile-up effect must be considered even in the observation mode described above. The standard method to mitigate pile-up in CCD image analysis is to exclude the central region of the image when extracting events. Based on the empirical criteria of \citet{Yamada2012-qx}, with modifications for Xtend, we estimate a pile-up fraction of $\sim$3\% within a 20$''$ radius ($\sim$10 pixels). We extracted the source events from an annular region centered on the point source, with an inner radius of 10 pixels and an outer radius of 50 pixels. The validity of this choice was confirmed by comparing spectra extracted from different regions, where no statistically significant differences beyond the expected uncertainties were observed (see Appendix 1). The RMF was generated using \texttt{xtdrmf}. The exposure map was generated using \texttt{xaexpmap}, and the ARF was generated using \texttt{xaarfgen} for the annular region. We did not subtract the non-X-ray background (NXB) in this analysis, as it remained below 1\% of the source signal across the entire energy band.

\section{Data Analysis and Results}\label{sec:data analysis and results}
\subsection{Light Curves and Dip Selection}\label{subsec:Light Curves and Dip Selection}
\begin{figure*}[h]
  \begin{center}
    \includegraphics[width=\linewidth]{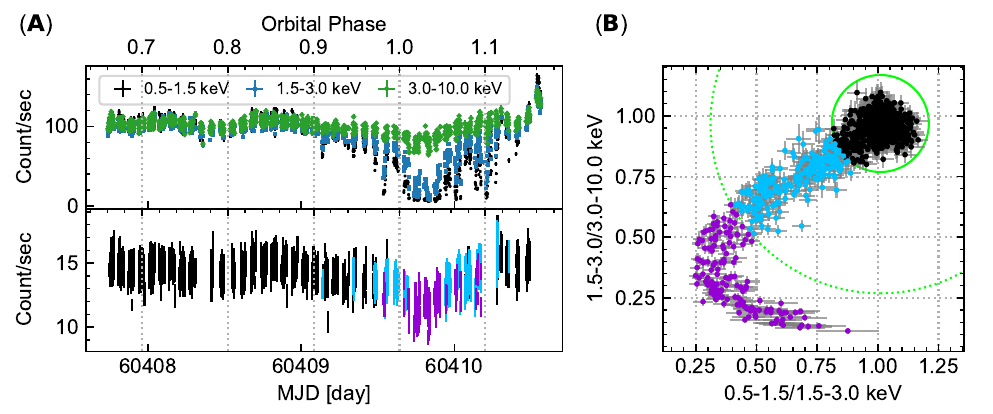}
  \end{center}
  \caption{(A) Xtend light curves (upper) in 0.5--1.5 (black), 1.5--3.0 (blue), and 3.0--10.0~keV (green), and the Resolve light curve (lower) in 2.0--10.0~keV. The bin size was set to 100~s. Times recorded by Xtend and Resolve are converted to MJD (first axis) and then to orbital phase (second axis). The 0.5--3.0~keV light curve shows a significant reduction in intensity corresponding to the X-ray absorption dip at superior conjunction ($\phi_{\mathrm{orb}}\approx0$). In the Resolve light curve (lower panel), the cyan and purple data points indicate the weak-dip and strong-dip intervals, respectively, as defined in Section~\ref{subsec:Light Curves and Dip Selection}. (B) Color-color diagram. Each color ratio (0.5--1.5~keV/1.5--3.0~keV, 1.5--3.0~keV/3.0--10.0~keV) was calculated from the light curves shown in Figure \ref{fig:lc_and_cc_d}(A). Two circles with radii of 0.2 (lime, solid) and 0.75 (lime, dashed) were defined, centered on the point with the highest counts in the 2D histogram created from the scatter plot. We defined non-dip, weak-dip (cyan), and strong-dip (purple) regions from the inside outward, with each circle as a boundary. \\
{Alt text:Panel (A) consists of two vertically stacked sub-panels, each showing Xtend and Resolve light curves. Panel (B) displays color-color diagram constructed from two X-ray hardness ratios: the horizontal axis represents the ratio of counts in the 0.--1.5 keV and 1.5--3.0 keV bands, and the vertical axis represents the ratio of counts in the 1.5--3.0 keV and 3.0--10.0 keV bands.}
  }
  \label{fig:lc_and_cc_d}
\end{figure*}

We created light curves in three energy bands (0.5--1.5, 1.5--3.0, and 3.0--10.0~keV), as shown in Figure \ref{fig:lc_and_cc_d}(A). The light curves were generated with a time bin size of 100~s and incorporated the effective exposure time, accounting for the lost time from the good time intervals (GTIs) of the cleaned events. The orbital phases of Cyg~X-1 were calculated from the Xtend observation times, adopting the orbital elements from \citet{Brocksopp1999-zn}. 

The light curves exhibit a clear dependence on orbital phase. In the 0.5--3.0~keV band, the count rate begins to decrease gradually from orbital phase $\phi_{\mathrm{orb}}\approx0.95$ and then steadily recovers after $\phi_{\mathrm{orb}}\approx0.05$, corresponding to a strong absorption dip. In addition to this overall trend, the soft band also shows short-term decreases in brightness at phases 0.91 and 0.98, corresponding to short dips \citep{Kitamoto1984-hc}. In contrast, in the 3.0--10.0~keV band, the count rate remains nearly constant throughout the entire observation. At orbital phases $\phiorb \sim $0.02--0.04, an increase in the 1.5--3.0~keV count rate leads to a change in hardness. This characteristic behavior has also been reported in previous studies (e.g., \cite{Hirsch2019-em}). 

We employed a classification method using a color-color diagram to distinguish between dip and non-dip events \citep{Nowak2011, Hirsch2019-em}. The dips are caused by absorption of soft X-rays by material crossing the line of sight to the X-ray source. Variations of spectra and color-color diagrams as a function of column density were analyzed in detail by \citet{Hirsch2019-em}. We created the color-color diagram shown in Figure \ref{fig:lc_and_cc_d}(B) using the light curves for the three energy ranges in Figure \ref{fig:lc_and_cc_d}(A). The overall shape of the color-color diagram is consistent with previous reports, indicating that the X-ray emission from Cyg~X-1 is absorbed by gas along the line of sight. In particular, the orbital-phase-dependent track suggests that this behavior is caused by local gas around Cyg~X-1. A $50\times50$ two-dimensional histogram was constructed from the scatter plot in Figure \ref{fig:lc_and_cc_d}(B). We identified the bin with the highest frequency (i.e., the point of maximum density) and set a circle of radius 0.2 centered on that point. The events within this circle were defined as non-dip periods, and those outside as dip phases. For the dip phase, we defined a circle of radius 0.75 to divide the dip phase into a weak-dip (inner region) and a strong-dip (outer region), and extracted the corresponding events. The dip phase is concentrated at $\phiorb$ 0.98--0.1, particularly at 0.02--0.08 for the strong-dip (shown in Figure \ref{fig:lc_and_cc_d}(A)).

\subsection{Modeling of Fe Emission and Absorption Lines}\label{subsec:Modeling of Fe Emission and Absorption Lines}

In this section, we explore a baseline model for quantitative line fitting to evaluate the emission- and absorption-line parameters observed with Resolve. First, for the emission line, we examined model fits to the non-dip spectrum in the 6.2--6.6~keV band. The models considered were (A) a single-Gaussian profile, (B) a double-Gaussian profile, (C) a double-Lorentzian profile, and (D) the H\"olzer model. In all cases, the line-center energies of all Gaussian, Lorentzian, and Voigt components were fixed to their respective transition energies, and any required energy shift was implemented via a redshift parameter applied to the entire model. For Model (A), a single Gaussian component was used with its line-center energy fixed at 6.4~keV. For Model (B), two Gaussian components were employed, with line-center energies fixed at the Fe\,I K$\alpha_{1}$ ($2p_{3/2}\rightarrow1s$) transition at 6.404~keV and the Fe\,I K$\alpha_{2}$ ($2p_{1/2}\rightarrow1s$) transition at 6.393~keV. The line widths $\sigma$ were set to a common value. In accordance with quantum-mechanical probabilities, the line-intensity ratio was fixed at 2:1 (K$\alpha_{1}$:K$\alpha_{2}$). Model (C) was constructed from Model (B) by substituting Lorentzian functions for the Gaussian functions. Model (D) is a phenomenological model that describes the Fe K$\alpha$ line spectrum using seven Voigt functions. We fixed the line-center energies and intensity ratios of the seven Voigt components according to \citet{Holzer1997-cr}. For each model, we compared four configurations in which the Gaussian line width, $\sigma$, and redshift, $z_G$, were either fixed at zero or left free (the line flux remained free in all cases). 

\begin{figure*}[h]
  \begin{center}
    \includegraphics[width=\linewidth]{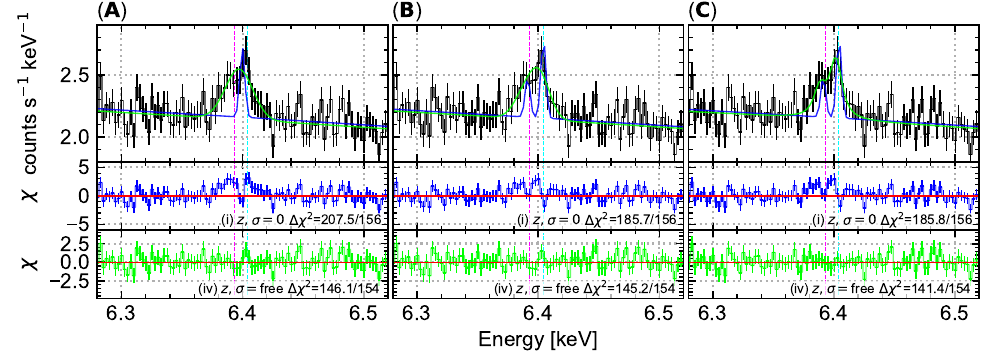}
  \end{center}
  \caption{Comparison of four models for the Fe I K$\alpha$ emission line: (A) single-Gaussian, (B) double-Gaussian, (C) double-Lorentzian. For each model, residuals from the four configurations with line width $\sigma$ and redshift $z$ are shown in the second and subsequent rows of panels in different colors ((i):blue, (iv):lime). In the top panel, the model for each configuration is displayed in each color. The dotted lines in magenta and cyan indicate the energies of Fe I K$\alpha_{2}$ at 6.393~keV and Fe I K$\alpha_{1}$ at 6.404~keV, respectively.\\
{Alt text: Each of the four panels shows the spectrum with several model curves in the top plot and four residual panels below. }
  }
  \label{fig:ResolveFeKa_linemodel}
\end{figure*}

In Figure \ref{fig:ResolveFeKa_linemodel}, we present the results for Models (A), (B), and (C) under two representative configurations out of the four tested: (i) $z_G$ = 0 and $\sigma$ = 0 fixed, and (iv) both $z_G$ and $\sigma$ treated as free parameters. Equivalent widths calculated from the $z$- and $\sigma$-free fits in Models (A) and (B) yield values of 3.5~eV. The observed line profile has a complex asymmetric shape and is not well reproduced when represented by a Gaussian width $\sigma$. In Model (C), introducing a Lorentzian width $\Gamma$ enables the K$\alpha_{1}$ and K$\alpha_{2}$ profiles to be reproduced, but the fit indicates overfitting ($\Delta \chi^2 = 141.1/154$). From a statistical perspective, the best fit is obtained with Model (D)(i) $z=0, \sigma=0$. While it is difficult to assess the quality of each model solely through statistical criteria, it is evident that the line profile exhibits an asymmetric shape due to the superposition of K$\alpha_{2}$ and K$\alpha_{1}$ lines with an intensity ratio of 2:1, which cannot be fully represented by these approximations. Detailed results are summarized in Table \ref{tab:Fitting FeKa}.

\begin{table*}[htb]
    \tbl{Results of fits to the averaged Fe K$\alpha$ profile using four emission models.}{%
    \begin{tabular}{ccccccccc}
    \hline \hline
      ~ & (A) & ~ & ~ & ~ & (B) & ~ & ~ &   \\ 
      ~ & (i) & (ii) & (iii) & (iv) & (i) & (ii) & (iii) & (iv) \\ \hline
      $\sigma$~[eV] & - & - & 11.1$_{-1.6}^{+1.8}$ & 11.3$_{-1.4}^{+1.7}$ & - & - & 8.6$\pm$ 2.0 & 8.2$_{-3.0}^{+2.0}$  \\ 
      $z(\times10^{-4})$ & - & $-3.7_{-1.2}^{+0.8}$ & - & $4.6_{-2.0}^{+3.0}$ & - & 2.9$_{-0.5}^{+1.0}$ & - & 4.5$\pm$ 2.0  \\ 
      $\chi^2$/dof & 207.5/156.0 & 196.2/155.0 & 149.8/155.0 & 146.1/154.0 & 185.7/156.0 & 173.7/155.0 & 149.4/155.0 & 145.2/154.0  \\ 
      \hline 
      ~ & (C) & ~ & ~ & ~ & (D) & ~ & ~ &   \\ 
      ~ & (i) & (ii) & (iii) & (iv) & (i) & (ii) & (iii) & (iv) \\ \hline
      $\sigma$~[eV] & - & - & 12.2$_{-4.0}^{+5.0}$ & 10.2$_{-3.0}^{+4.0}$ & - & - & 7.3$_{-4.0}^{+3.0}$ & 7.2$_{-4.0}^{+3.0}$  \\ 
      $z(\times10^{-4})$ & - & 2.9$_{-0.6}^{+1.1}$ & - & 3.6$_{-1.5}^{+1.7}$ & - & 1.9$\pm$ 1.0 & - & 3.2$\pm$ 2.0  \\ 
      $\chi^2$/dof & 185.8/156.0 & 173.8/155.0 & 146.7/155.0 & 141.4/154.0 & 157.3/156.0 & 153.2/156.0 & 147.2/155.0 &  \\
    \hline\hline
    \end{tabular}
    }
    \label{tab:Fitting FeKa}
\end{table*}

In this paper, we employ a photoionized absorption model to reproduce the observed absorption lines. Two model sets are used: \Cloudyabs, generated with the Cloudy code \citep{Chatzikos2023-ym,Gunasekera2023-vr}, and \texttt{pion} (hereafter \pionabs), the photoionized plasma model in SPEX. The \Cloudyabs\ model used in this paper is the same as that used in Y25. For each model, the spectral energy distribution (SED) used to photoionize the gas was constructed from the parameters of the Xtend and NuSTAR spectra, obtained by fitting with \texttt{tbabs$\times$(diskbb+nthComp)} (discribed in Appendix \ref{Appendix:sec:SED}). The parameters were $T_{\text{in}} = 0.166$~keV and norm = $6.11\times10^5$ for \texttt{diskbb}; $\Gamma = 1.67$, $kT_e = 100$~keV, $kT_{\text{bb}} = 0.166$~keV, inp\_type = 0, redshift = 0, and norm = 1.50 for \texttt{nthComp}. The incident SED used for Cloudy is shown in Appendix Figure \ref{fig:appendix_cloudy_sed}. In addition, the phenomenological ion absorption model \texttt{Ionabs}, created with the Voigt profile and ion database, was also applied (the same as the \texttt{kabs} model; \cite{Ueda2004-tv}). Since one \texttt{Ionabs} component corresponds to a single ion, we included two ions, Fe\,\textsc{xxv} and Fe\,\textsc{xxvi}, for which Doppler broadening and redshift were tied together. 

These models share four common free parameters: column density $N({\mathrm{H}})$, redshift $z$, turbulent velocity $v_{\mathrm{turb}}$, and ionization parameter $\xi$. We calculated the Doppler broadening $v_{\mathrm{D}}$ from $v_{\mathrm{turb}}$ assuming a gas temperature of $T = 1.0\times10^5$~K and a mass number of 56 ($9.3\times10^{-26}$~kg). Table~\ref{tab:fitting results of three absmodel} summarizes the fit results. All three models yield consistent values for $z$ and $v_{\mathrm{D}}$. Differences in $\log \xi$ and $N({\mathrm{H}})$ between \Cloudyabs\ and \pionabs\ arise from the adopted elemental abundance sets and the details of the ionization/radiation calculations. Although the absolute values of $\log \xi$ and $N({\mathrm{H}})$ differ systematically across models, the resulting absorption-line profiles are essentially identical. For reproducibility and consistency, we adopt the \Cloudyabs\ model to evaluate the absorption-line parameters.

\begin{table}
    \tbl{Results of fits to the spectra with absorption lines using three models.}{%
    \begin{tabular}{cccc}
        \hline\hline
        ~ & \texttt{Ionabs} & \texttt{pion} & \texttt{Cloudyabs}  \\ \hline
        $\log \xi$ & - & 3.09$_{-0.04}^{+0.03}$ & 3.12$\pm$ 0.03  \\ 
        $z/10^{-4}$ & $-2.7\pm0.5$ & $-2.4\pm0.5$ & $-2.9\pm0.5$  \\ 
        $N(\text{H})/10^{22} [\mathrm{cm}^{-2}]$ & 0.546$\pm$ 0.06 & 0.775$_{-0.08}^{+0.09}$ & 0.9$_{-0.1}^{+0.11}$  \\ 
        $v_{\text{D}} [km]$ & 148.6$_{-24.0}^{+29.0}$ & 114.1$_{-42.0}^{+26.0}$ & 140.4$_{-27.0}^{+28.0}$  \\ 
        \hline
        $\chi^2$/dof & 1535.6/1393 & 1518.5/1393 & 1523.4/1393 \\ \hline\hline
    \end{tabular}
    }
    \label{tab:fitting results of three absmodel}
\end{table}


\subsection{Dip vs.\ Non-dip Spectral Variability}\label{subsec:Spectral Variability between Non-dip and Dip}

\begin{figure*}[h]
  \begin{center}
    \includegraphics[width=\linewidth]{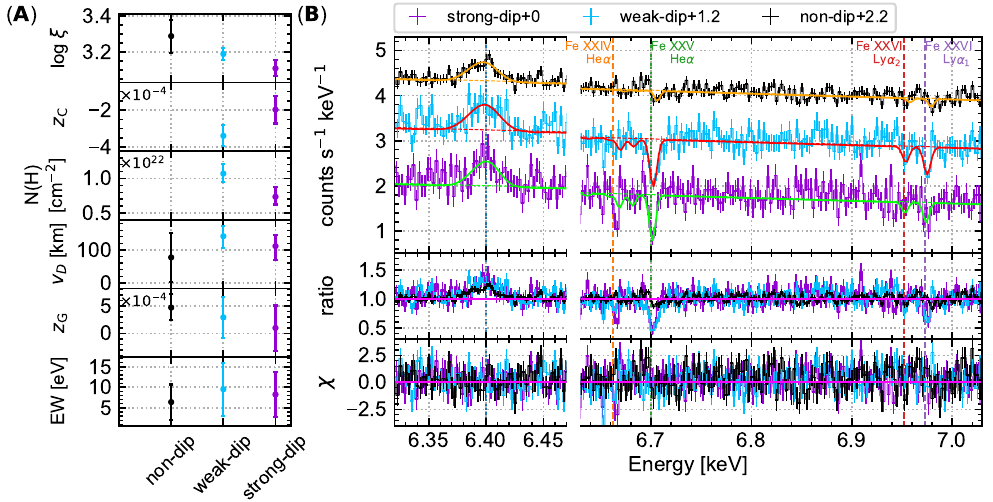}
  \end{center}
  \caption{Results of the \Cloudyabs~and double Gaussian model fits to the Resolve spectra at each dip level. (A) The parameter for each dip level. (B) The spectra and residuals for each dip level and models. The spectral fit was performed over the 6.0--10.0~keV, and this figure presents a zoomed-in view of the 6.3--6.48~keV and 6.63--7.02~keV. (Top)Resolve spectra at each dip level, plotted with arbitrary offsets (non-dip: black, weak-dip: cyan, strong-dip: purple), with the corresponding best-fit model given by \Cloudyabs~and double Gaussian. (Middle) Ratios to \texttt{powerlaw} component for each dip level. (Bottom) Residuals of the spectra at each dip level fitted with \Cloudyabs~and double Gaussian.
{Alt text: The figure is composed of two panels. Panel (A) shows a set of parameter plots comparing non-dip, weak-dip, and strong-dip intervals. Each vertical series of points with error bars represents model parameters. Panel (B) displays spectra (upper), ratios (middle) and residuals with best-fit model (lower) for the three dip levels.}
  }
  \label{fig:nondip_dip_spec}
\end{figure*}

We extracted the non-dip, weak-dip, and strong-dip spectra of Resolve using the time intervals defined in Section~\ref{subsec:Light Curves and Dip Selection}, as shown in Figure~\ref{fig:nondip_dip_spec} (top). The effective exposures for each period were 79~ks, 19~ks, and 14~ks, respectively. We analyzed the spectra in the 6.0--10.0~keV band.
Figure~\ref{fig:nondip_dip_spec} (middle) shows residuals to a power-law fit. The power-law model roughly describes the overall continuum. The photon index $\Gamma$ is $\sim$1.7 in the non-dip spectrum and $\sim$1.4 in the dip spectrum. From the spectra and the residuals relative to the power-law continuum, the Fe\,\textsc{xxv} He$\alpha$ ($1s^{2}\,{}^{1}\!S_{0} \rightarrow 1s2p\,{}^{1}\!P_{1}$) absorption at 6.700~keV and the Fe\,\textsc{xxvi} Ly$\alpha_{2}$ ($1s\,{}^{2}\!S_{1/2} \rightarrow 2p\,{}^{2}\!P_{1/2}$) and Ly$\alpha_{1}$ ($1s\,{}^{2}\!S_{1/2} \rightarrow 2p\,{}^{2}\!P_{3/2}$) absorptions at 6.952~keV and 6.973~keV, respectively, are clearly detected. 

We fitted the 6.0--10.0~keV spectra with the model \texttt{tbabs$\times$(zashift$\times$(gauss+gauss) + \Cloudyabs$\times$powerlaw)}. Because the effect of interstellar absorption does not significantly impact the results of fitting with the 6.0--10.0~keV band, we fixed the hydrogen column density of \texttt{tbabs} at $6.0\times10^{21}~\text{cm}^{-2}$ \citep{Makishima2008-wn}. 

Given the limited spectral quality, the Fe\,I K$\alpha$ emission line was modeled with a double Gaussian (a practical simplification that captures the observed profile without implying detailed physical modeling). The absorption lines were modeled with \Cloudyabs. The fits are shown in Figure~\ref{fig:nondip_dip_spec} (bottom) and summarized in Table~\ref{tab:fitting results of each dip level}. The parameters of the Fe\,I K$\alpha$ emission line remain unchanged within the sensitivity of the present analysis. In contrast, the absorption lines exhibit clear variability. During the non-dip period, the He-like Fe line is weak, corresponding to a column density of $(0.23^{+0.08}_{-0.07})\times10^{22}~\mathrm{cm}^{-2}$, a redshift of $(-9.1\pm1.2)\times10^{-4}$, and an ionization parameter $\log \xi = 3.28^{+0.09}_{-0.08}$. Comparing the weak-dip and strong-dip spectra, the strong-dip shows lower ionization (smaller $\xi$), a reduced blueshift (the \Cloudyabs\ redshift $z_{\mathrm{C}}$ closer to zero), a lower column density $N(\mathrm{H})$, and a smaller velocity dispersion (smaller Doppler broadening).

A Li-like iron (Fe\,\textsc{xxiv}) K-line at 6.662~keV ($1s^{2}2s\,{}^{2}\!S_{1/2} \rightarrow 1s2s2p\,{}^{2}\!P_{3/2}$) is clearly present in the strong-dip spectrum. This feature -- previously identified in laboratory measurements (e.g., \cite{Rudolph2013-vu}) -- cannot be reproduced by a single-zone photoionized absorber tuned to match the He-like (Fe\,\textsc{xxv}) and H-like (Fe\,\textsc{xxvi}) lines. We therefore fitted the spectrum with three independent \texttt{Ionabs} components representing the Li-like (Fe\,\textsc{xxiv}), He-like (Fe\,\textsc{xxv}), and H-like (Fe\,\textsc{xxvi}) ions, and summarize the results in Table \ref{tab:fitting results of each dip level}. In the strong-dip, the Fe\,\textsc{xxvi} ion column density decreases whereas the Fe\,\textsc{xxv} ion column density increases relative to the weak-dip, indicating a lower ionization state. The line centroids also imply different bulk velocities: Fe\,\textsc{xxv} has $z = (-1.8^{+0.8}_{-0.7})\times10^{-4}$, while the Li-like Fe\,\textsc{xxiv} component shows a larger blueshift, $z = (-5.4^{+1.6}_{-1.4})\times10^{-4}$. These results point to multi-zone (or clumpy) absorbing material with distinct ionization and kinematic properties along the line of sight. A full two-zone photoionization treatment is deferred and will be reported in a subsequent paper.

\begin{table}[h]
    \tbl{Best-fit parameters for the non-dip, weak-dip, and strong-dip spectra with \Cloudyabs$\times$\texttt{powerlaw} $+$ Fe\,K$\alpha$ emission and three independent \texttt{Ionabs} for H-, He-, Li-like Fe.}{%
    \begin{tabular}{cccc}
    \hline
    \hline
      ~ & Non-dip & Weak-dip & Strong-dip  \\ \hline
      $\log \xi$ & 3.288$\pm$ 0.09 & 3.189$\pm$ 0.03 & 3.109$_{-0.04}^{+0.05}$  \\ 
      $z_{\mathrm{C}}/10^{-4}$ & $-9.1\pm$ 1.3 & $-3.38\pm$ 0.6 & $-1.97\pm0.8$  \\ 
      $N$(H)$/10^{22}$ [cm$^{-2}$] & 0.2$_{-0.06}^{+0.09}$ & 1.07$_{-0.12}^{+0.14}$ & 0.73$_{-0.11}^{+0.15}$  \\ 
      $v_{\mathrm{D}}$ [km] & 77.4$_{-76.0}^{+73.0}$ & 141.8$_{-37.0}^{+32.0}$ & 111.7$_{-42.0}^{+33.0}$  \\ 
      Photon index & 1.985$\pm$ 0.012 & 1.978$\pm$ 0.03 & 2.01$\pm$ 0.03  \\ 
      $z$$_{\mathrm{G}}/10^{-4}$ & 4.7$\pm$ 2.0 & 2.9$\pm$ 4.0 & 1.0$\pm$ 4.0  \\ 
      EW [eV] & 6.4$\pm$ 4.0 & 9.5$\pm$ 6.0 & 8.2$\pm$ 6.0  \\ 
      \hdashline
      $\chi^2$/dof & 1818.0/1590 & 1718.7/1590 & 1658.7/1590  \\ 
      \hline
      $z_{\text{Fe\,\textsc{xxvi}}}/10^{-4}$ & - & $-2.5\pm0.7$ & $-3.0\pm 1.9$  \\ 
      $N_{\text{Fe\,\textsc{xxvi}}}/10^{18} [\text{cm}^{-2}]$ & - & 0.16$_{-0.05}^{+0.20}$ & 0.063$\pm$ 0.02  \\ 
      $z_{\text{Fe\,\textsc{xxv}}}/10^{-4}$ & - & $-3.7\pm0.8$ & $-1.8_{-0.7}^{+0.8}$  \\ 
      $N_{\text{Fe\,\textsc{xxv}}}/10^{18} [\text{cm}^{-2}]$ & - & 0.105$_{-0.012}^{+0.013}$ & 0.15$_{-0.06}^{+0.08}$  \\ 
      $z_{\text{Fe\,\textsc{xxiv}}}/10^{-4}$ & - & $-2.46_{-0.05}^{+0.06}$ & $-5.4_{-1.4}^{+1.6}$ \\ 
      $N_{\text{Fe\,\textsc{xxiv}}}/10^{18} [\text{cm}^{-2}]$ & - & 0.086$_{-0.05}^{+0.06}$ & 0.036$_{-0.013}^{+0.015}$ \\ 
    \hline
    \hline
    \end{tabular}
    }
    \label{tab:fitting results of each dip level}
\end{table}

\begin{figure}[h]
  \begin{center}
    \includegraphics[width=\linewidth]{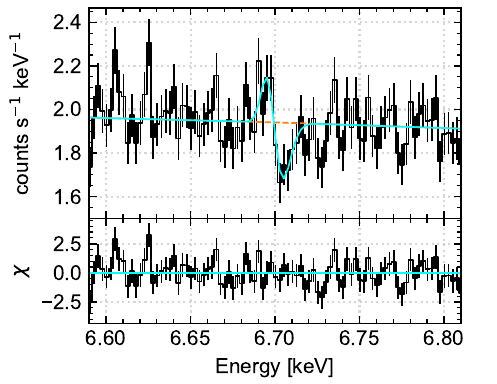}
  \end{center}
  \caption{
  P~Cygni-like profile of the Fe\,\textsc{xxv} He$\alpha$ (6.7~keV) line in the non-dip spectrum. The bottom panel shows the fit with two Gaussian components (emission and absorption) with independent redshifts.
{Alt text: The figure consists of two stacked panels, showing the spectrum around 6.7 keV in the upper panel and the residuals in the lower panel.}
  }
  \label{fig:PCygni_in_non-dip}
\end{figure}

The non-dip spectrum exhibits both absorption and emission features at 6.7~keV from He-like iron (Fe\,\textsc{xxv}), forming a P~Cygni-like profile. We fitted the feature with two Gaussian components, one for the emission line and one for the absorption line, each with an independent redshift $z$. Figure~\ref{fig:PCygni_in_non-dip} shows a redshifted emission line at $z=6.1\times10^{-4}$ and a blueshifted absorption line at $z=-2.7\times10^{-4}$. Fixing the photon index and the normalization of the power-law continuum to the best-fit values, the equivalent widths of the emission and absorption components are $(2.0^{+2.0}_{-1.2})$~eV and $(2.5^{+7.0}_{-1.3})$~eV, respectively. 
Taken together, the P~Cygni-like profile in the non-dip spectrum suggests outflowing, line-of-sight absorbing gas and a redshifted, possibly inflowing emitting component. However, the detailed interpretation depends sensitively on the assumed stellar-wind geometry and radiative transfer, and a self-consistent model is beyond the scope of this paper. We therefore confine ourselves here to reporting the observational evidence for a P~Cygni-like structure.

\subsection{Orbital Phase-Resolved Spectroscopy}

\begin{figure}[h]
  \begin{center}
    \includegraphics[width=\linewidth]{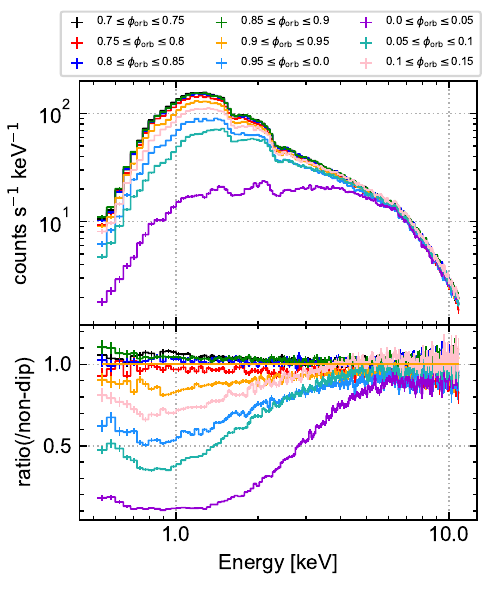}
  \end{center}
  \caption{
Orbital-phase-resolved spectra of Xtend and their ratios to the non-dip spectrum. (Top) Detector-response-folded Xtend spectra. As indicated in the legend, each color corresponds to an orbital phase bin ($\Delta\phi_{\mathrm{orb}}=0.05$). (Bottom) Ratios to the non-dip spectrum defined in Section~\ref{subsec:Light Curves and Dip Selection}, computed using the detector-response-folded spectra. The ratio shows a marked change below $\sim$3~keV; the spectrum at $0.00\le\phi_{\mathrm{orb}}\le0.05$ is $\sim$20\% lower in intensity than the non-dip spectrum.
{Alt text: The figure consists of two stacked panels. The upper panel shows multiple spectra across different orbital phase bins, distinguished by color. The lower panel shows the ratios of these spectra to the non-dip spectrum.}
} \label{fig:OrbitalPhaseResolvedSpectra_xtd}
\end{figure}

To further examine the time dependence of the absorption structure, we analyzed orbital phase-resolved spectra. We extracted spectra by sliding orbital-phase windows of width $\Delta\phi_{\mathrm{orb}}=0.05$ (step 0.05) and $\Delta\phi_{\mathrm{orb}}=0.10$ (step 0.05). The orbital elements were adopted from \citet{Brocksopp1999-zn}. Figure~\ref{fig:OrbitalPhaseResolvedSpectra_xtd} shows the Xtend orbital phase--resolved spectra and their ratios to the non-dip spectrum. Spectral changes are concentrated below $\sim3$~keV, with the strongest absorption at $0.00\le\phi_{\mathrm{orb}}\le0.05$. In this phase range, the 0.5--2.0~keV flux is $\sim$20\% of the non-dip level. Compared with the adjacent phase intervals ($0.95\le\phi_{\mathrm{orb}}\le1.00$ and $0.05\le\phi_{\mathrm{orb}}\le0.10$), the variability occurs on short timescales set by $\Delta\phi_{\mathrm{orb}}\approx0.05$--0.10 (i.e., a few to $\sim$10~hours for $P_{\mathrm{orb}}=5.59$~d). Additionally, in these phases the ratio is energy dependent: the flux is reduced by approximately 50\% at 0.1~keV, 40\% at 0.9~keV, and 50\% at 1.5~keV.

\begin{figure*}[h]
  \begin{center}
    \includegraphics[width=0.85\linewidth]{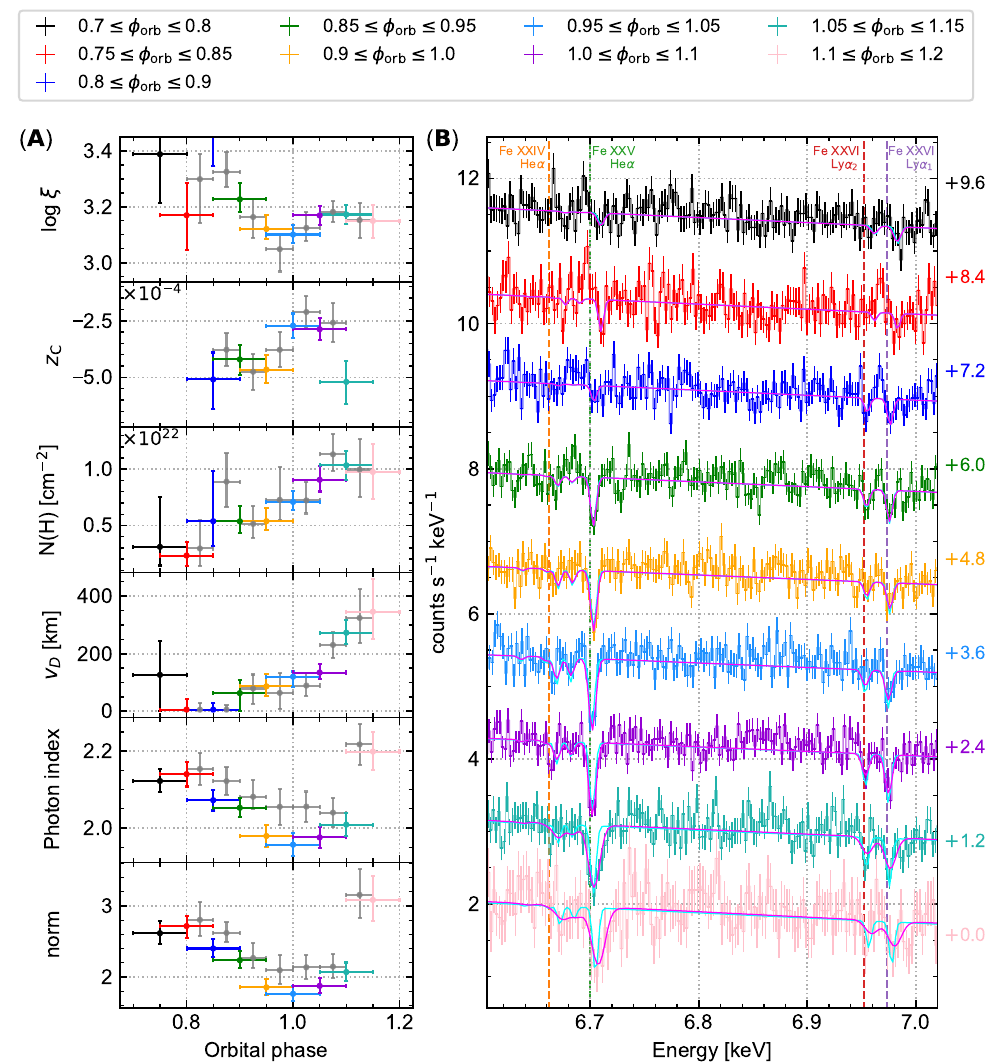}
  \end{center}
  \caption{
(A) Absorption parameters as a function of orbital phase. As indicated in the legend, colored points show results for $\Delta\phi_{\mathrm{orb}}=0.10$, while gray points show results for $\Delta\phi_{\mathrm{orb}}=0.05$ slid in steps of 0.05. The parameters $\log\xi$, $N(\mathrm{H})$, and $z$ derived from the Fe\,\textsc{xxv}/Fe\,\textsc{xxvi} absorption show clear orbital modulation. (B) Orbital-phase-resolved Resolve spectra for each phase bin (legend). The magenta solid curve is the best-fit model; the cyan solid curve shows the model with the line width ($v_{\mathrm{D}}$) fixed at zero. Each spectrum is vertically offset by a constant for clarity. The orange, green, and red dotted lines mark the rest energies of Fe\,\textsc{xxv} He$\alpha$, Fe\,\textsc{xxvi} Ly$\alpha_{1}$, and Fe\,\textsc{xxvi} Ly$\alpha_{2}$, respectively.
{Alt text: The figure consists of two panels. Panel (A) shows best-fit parameters plotted against orbital phase, with data points in different colors representing phase bins. Panel (B) displays spectra for each orbital phase, vertically offset for clarity, with best-fit model curves and vertical dashed lines marking the positions of key Fe absorption lines.}
}
  \label{fig:OrbitalPhaseResolvedSpectra_rsl}
\end{figure*}

We next analyzed absorption above 6.5~keV using Resolve spectra (6.5--10.0~keV). Each spectrum was fitted with \texttt{tbabs}($nH=6.0\times10^{22}$)$\times$\Cloudyabs$\times$\texttt{powerlaw}. We show the results in Figure \ref{fig:OrbitalPhaseResolvedSpectra_rsl} We find a clear orbital-phase dependence in both the continuum and the absorption parameters. The ionization parameter varies gradually with phase, reaching a minimum near superior conjunction ($0.95\le\phi_{\mathrm{orb}}\le1.00$ and $0.00\le\phi_{\mathrm{orb}}\le0.05$). The redshift is $\sim-5\times10^{-5}$ for $\phi_{\mathrm{orb}}<0$ and then shifts abruptly to $\sim-2.5\times10^{-5}$ for $0.00\le\phi_{\mathrm{orb}}\le0.10$. In terms of radial velocity, this corresponds to a change from $\sim150~\mathrm{km\,s^{-1}}$ to $\sim75~\mathrm{km\,s^{-1}}$. For $\phi_{\mathrm{orb}}>0.0$, the Fe\,\textsc{xxv} He$\alpha$ absorption line tends to broaden (larger Doppler width). Overall, the phase-resolved behavior is consistent with the sightline intersecting a focused, clumpy wind near superior conjunction. The evolution of $\log \xi$ and $z$, together with tentative line broadening, suggests multiple kinematic components. A self-consistent wind model is beyond the scope of this paper and will be presented in a subsequent publication.

\subsection{Intensity-sorted Spectroscopy}

In this study, we further applied the ``intensity-sorted spectroscopy'' technique, originally introduced by \citet{Makishima2008-wn}, to investigate possible short-timescale (a few seconds) variability. \citet{Makishima2008-wn} demonstrated the method using a Suzaku observation of Cyg~X-1, showing that spectral changes can be tracked as a function of source intensity. Subsequently, \citet{Yamada2013-wg} performed a systematic analysis of 25 hard-state Suzaku observations of Cyg~X-1 (excluding dip phases) and confirmed that this approach effectively probes variability on second timescales.

We performed intensity sorting using the Resolve light curve in the dip phase defined in Section~\ref{subsec:Light Curves and Dip Selection}. The light curve was constructed from events of all grades in the 2.0--10.0~keV band. We defined the high-flux phase as
\begin{equation}
  \left\{\, t \,\middle|\, C(t) \ge \overline{C(t)}_{\Delta T} + f_{\mathrm{h}}\,\sigma(t) \,\right\},
  \label{eq:fh}
\end{equation}
and the low-flux phase as
\begin{equation}
  \left\{\, t \,\middle|\, C(t) < \overline{C(t)}_{\Delta T} - f_{\mathrm{l}}\,\sigma(t) \,\right\}.
  \label{eq:fl}
\end{equation}
Here, $C(t)$ and $\sigma(t)$ are the count rate and its Poisson error at time $t$, respectively. The parameters $f_{\mathrm{h}}$ and $f_{\mathrm{l}}$ are the threshold factors for the high- and low-flux selections. The running mean $\overline{C(t)}_{\Delta T}$ is the average count rate over $\Delta T$ seconds, evaluated for each segment by dividing the dip interval into consecutive $\Delta T$-second segments starting from its beginning. Based on Eqs.~(\ref{eq:fh}) and (\ref{eq:fl}), time bins were classified as high- or low-flux phases, and the remaining bins were assigned to the middle-flux phase.

The free parameters are the light-curve bin size $\delta t$, the segment length $\Delta T$, and the threshold factors $f_{\mathrm{h}}$ and $f_{\mathrm{l}}$. The bin size $\delta t$ should resolve brightness variations clearly; $\Delta T$ must be long enough to be insensitive to second-scale fluctuations yet short enough to follow global changes; and $f_{\mathrm{h}}$, $f_{\mathrm{l}}$ control the separation of intensity levels with a trade-off in photon statistics. After testing various configurations, we adopted $\delta t = 1.5$~s, $\Delta T = 100$~s, and $f_{\mathrm{h}} = f_{\mathrm{l}} = 1.0$.

\begin{figure*}[h]
  \begin{center}
    \includegraphics[width=\linewidth]{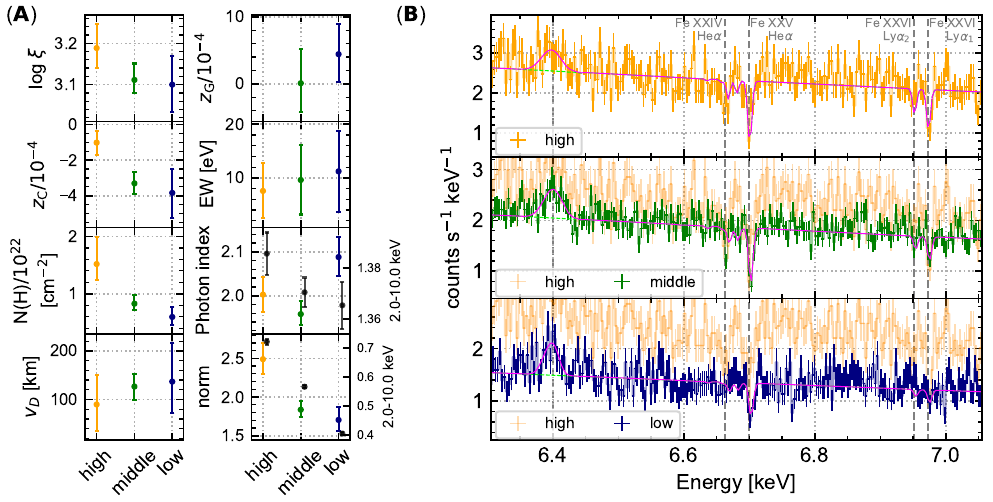}
  \end{center}
  \caption{
  (A) Results of the \Cloudyabs$\times$\texttt{powerlaw} + \texttt{double-Gaussian} fits. The continuum is modeled with a power law absorbed by \Cloudyabs; the double-Gaussian consists of two components for Fe\,I K$\alpha_{1}$ and Fe\,I K$\alpha_{2}$ sharing the same redshift and Gaussian width parameter. In the photon-index and normalization panels, gray points show the results of fitting the 2.0--10.0~keV spectra with a simple power-law model. (B) Spectra and best-fit models for the high- (top, orange), middle- (middle, green), and low-flux (bottom, blue) phases. The solid magenta curve shows the best-fit model, and the solid lime curve shows the \texttt{powerlaw} component.  
{Alt text: The figure consists of two panels. Panel (A) shows best-fit parameters plotted for the high-, middle-, and low-flux phases. Panel (B) displays the corresponding spectra and best-fit model curves for the same three flux phases, with vertical dashed lines marking the main Fe emission and absorption features.}
}
  \label{fig:IntesnsitySortedAna_spec}
\end{figure*}

Hp-grade spectra were extracted for the high-, middle-, and low-flux phases. When we first modeled the 2.0--10.0~keV band with a simple power law, the photon indices were found to be $1.385\pm0.008$, $1.370\pm0.006$, and $1.365\pm0.009$ for the high-, middle-, and low-flux phases, respectively (gray points in the photon-index panel of Figure~\ref{fig:IntesnsitySortedAna_spec}A). Thus, the spectrum softens as the source brightens, consistent with \citet{Makishima2008-wn} and \citet{Yamada2013-wg}. Following Section~\ref{subsec:Spectral Variability between Non-dip and Dip}, we then restricted the analysis to 6.0--10.0~keV and fitted \Cloudyabs\ for the absorption component, a double-Gaussian model for the emission lines, and power law for the continuum. The results are shown in Figure~\ref{fig:IntesnsitySortedAna_spec}. Toward lower flux, the ionization parameter decreases, the absorption lines become more blueshifted, the column density becomes smaller, and the inferred Doppler width tends to increase. For the emission-line component, no clear trend is found in the parameters (redshift $z_{\mathrm{L}}$ and equivalent width).

We varied the flux-selection parameters ($\delta t$, $\Delta T$, $f_{\mathrm{h}}$, and $f_{\mathrm{l}}$) and repeated the analysis. With $f_{\mathrm{h}}=f_{\mathrm{l}}=0$ (i.e., two classes: high and low only), we recovered the same trends between the absorption/Fe\,K$\alpha$ parameters and flux. For $f_{\mathrm{h}}$ and $f_{\mathrm{l}}>0.5$, the flux-dependent separation became clearer and spectral changes appeared larger; however, reduced statistics made it difficult to constrain the fine line-structure parameters. Balancing exposure and sensitivity to spectral changes, we adopted $\delta t = 1.5$~s.
Thus, the intensity-sorted results provide observational constraints on the ionization, kinematics, and structure of the focused, clumpy stellar wind and inform our understanding of its geometry, clumping, and radiative coupling to the X-ray source.

\section{Discussion}\label{sec:Discussion}

\subsection{Structured Wind Signatures}\label{subsec:Structured Wind Signatures}

XRISM observations have provided new understanding for a structured stellar wind in Cyg~X-1. The orbital-phase dependence of the light curves and absorption spectra obtained with Xtend (0.5--10.0~keV) and Resolve (6.0--7.0~keV) clearly reflects the properties of highly ionized wind material. The Xtend light curve below $\sim$3~keV is modulated with orbital phase and shows strong dimming at superior conjunction (Figure~\ref{fig:lc_and_cc_d}). The energy-dependent variability of the light curves is represented in the color-color diagram, and absorption is apparent around $\sim0.9$~keV in the spectra. These features are consistent with partial covering by gas with column densities of $N_{\mathrm{H}} \sim 10^{22}~\mathrm{cm}^{-2}$ \citep{Hirsch2019-em}. The tail-like structure of the color-color diagram and orbital-variability of the Xtend spectra implies an energy-dependent effect, which may include contributions from dust-scattering \citep{Nowak2011}. detailed characterization of the individual absorbing components is beyond the scope of this paper, which focuses on orbital-dependent variability, and is left for future work. In the Resolve 6.0--10.0~keV band, dip spectra are described by a power law plus a photoionized absorber: H-like (Fe\,\textsc{xxvi}) and He-like (Fe\,\textsc{xxv}) lines appear only during dips, and Li-like Fe\,\textsc{xxiv} emerges in the strong dips. This result provides evidence for the presence of highly ionized gas along the line of sight, and the orbital modulation of the absorption lines traces the orbital dependence of the interaction between the stellar wind and the primary X-ray radiation from the vicinity of the black hole.

In addition, the observed properties point to a stellar wind with complex kinematic and thermal structure. As shown in Figure~\ref{fig:nondip_dip_spec} and by the independent fits to Fe,\textsc{xxiv}/\textsc{xxv}/\textsc{xxvi}, the absorber exhibits non-uniform thermal and kinematic properties. In the \emph{non-dip} spectrum we also detect a P~Cygni-like feature at 6.7~keV (Fe\,\textsc{xxv} He$\alpha$), modeled with two Gaussians of independent redshift (Figure~\ref{fig:PCygni_in_non-dip}): a redshifted emission component at $z=6.1\times10^{-1}$ and a blueshifted absorption component at $z=-2.7\times10^{-4}$. The P~Cygni-like profile suggests outflowing, line-of-sight absorbing gas and a redshifted, possibly inflowing emitting component. 

These results agree with structured wind properties, photoionized-wind picture established by Chandra/HETG--deeper dips reveal lower-ionization species and complex Doppler shifts (\cite{Miskovicova2016-zi,Hirsch2019-em}). Our Resolve data extend this framework into the Fe\,K band: Fe\,\textsc{xxiv} is confined to the deepest dips, while Fe\,\textsc{xxv}/\textsc{xxvi} dominate typical dip intervals. The depth and duration of dips vary between orbits, consistent with the large scatter in $N({\mathrm{H}})$ at superior conjunction seen in long-term monitoring (\cite{Grinberg2015-kl}). The line-of-sight velocity of the absorber decreases markedly at deeper dip levels ($\phi \sim 0.0$--$0.1$), suggesting that the observed X-rays have passed through the region of the stellar wind converging toward the black hole before reaching the observer. Together with the non-dip P~Cygni-like signature, the evidence favors a structured, focused wind with embedded clumps and an asymmetric velocity field, rather than a smooth, steady outflow. 

\subsection{Origin of the Neutral Fe\,K$\alpha$ Line}

As shown in Figure~\ref{fig:ResolveFeKa_linemodel}, the average Fe\,\textsc{i} K$\alpha$ profile at 6.4~keV does not clearly separate the K$\alpha_1$ and K$\alpha_2$ components. Although Resolve’s $\sim$5~eV resolution should resolve the $\sim$13~eV doublet splitting, the inter-component valley is muted. Phase averaging can smear the line via orbit-induced Doppler shifts of a few eV, and the profile sharpness appears to vary with dip depth (Figure~\ref{fig:nondip_dip_spec}): the doublet is less apparent in deep dips and somewhat sharper in non-dip intervals. Given the limited S/N for time-resolved profiles, orbital Doppler smearing alone is unlikely to explain all observed variations.

Fluorescent emission from the stellar wind, reported in other HMXBs and often broader or complex, likely contributes in addition to disk reflection. A wind-origin component that varies with the line-of-sight wind column could blur the intrinsic doublet when spectra are averaged, naturally reducing the apparent K$\alpha_1$--K$\alpha_2$ separation. Disentangling disk and wind fluorescence will require phase-resolved spectroscopy with more highly statistics and radiative-transfer modeling in a structured wind. 

\subsection{Second-Scale Fe\,K Absorption Variability}
One of the most intriguing results from this observation is the tentative detection of variability in the highly ionized Fe\,K absorption features on timescales of only 1--2~s, correlated with the rapid X-ray intensity fluctuations (as shown in Figure \ref{fig:IntesnsitySortedAna_spec}). The significance is modest (of order $1\sigma$ for the line changes), but if confirmed this would be the first second-scale variability of wind absorption seen with microcalorimeter resolution, underscoring the unique capability of XRISM/Resolve to probe the stellar wind on timescales inaccessible to previous gratings.

The trends are consistent with ionization responding to flux. During brighter intervals, Fe\,\textsc{xxv} and Fe\,\textsc{xxvi} deepen and the highly ionized gas is more transparent above $\sim$6~keV, implying larger columns (or fractions) of H-/He-like Fe. In lower-flux intervals (second-scale dips/mini-dips), both $N({\mathrm{H}})$ and $\xi$ decrease while the lines become weaker but more blueshifted and broader (larger $v_{\mathrm{D}}$). This behavior can be explained by a cooler, denser and lower-ionized clump crossing the line of sight and reducing the ionizing field; spectral fits suggest a clump speed of $v\sim120~\mathrm{km\,s^{-1}}$, a few times that of the highly ionized absorber in bright intervals.

On 1-2 s timescales, the Fe\,K$\alpha$ emission shows no variability exceeding the 1$\sigma$ significance level. The inferred Fe\,K$\alpha$ parameters are consistent with only marginal flux-dependent trends, all consistent with statistical fluctuations.
While such behavior could in principle reflect delayed fluorescence or added wind fluorescence during continuum obscuration, the available statistics are insufficient to test these interpretations.

Overall, the second-scale spectral changes point to small-scale, clumpy structure and rapid wind-radiation coupling, consistent with the structured-wind picture in Section~\ref{subsec:Structured Wind Signatures}. They indicate that wind inhomogeneities can alter ionization and kinematics on very short spatial scales ($\sim10^{4}$--$10^{5}$~km).

Clumping further implies a lower effective mass-loss rate, helping to reconcile the $\sim$20--$21\,M_{\odot}$ black hole in Cyg~X-1 with near-solar metallicity environments \citep{Miller-Jones2021-nt}, with implications for massive-binary evolution and gravitational-wave progenitors. Longer, multi-orbit exposures will better establish second-scale variability and separate orbital from stochastic effects; this will enable more detailed discussion of wind clumping and line-driving processes. 

\section{Summary}\label{sec:Summary}
We carried out high-resolution X-ray spectroscopy of the black hole binary Cygnus~X-1 with XRISM during its Performance Verification phase. By combining the fine energy resolution of Resolve with the wide energy coverage of Xtend, we examined spectral variability as a function of orbital phase, focusing on both dip and non-dip phases around superior conjunction.

The properties of the highly ionized Fe absorption lines vary systematically with orbital phase. Changes in column density, ionization state, and blueshifted velocity indicate that the absorbing stellar wind is not uniform but contains structured components whose properties are influenced by the binary geometry. During deeper dips, the absorption spectrum shifts toward dominance by He-, H-like Fe with weaker blueshifts, suggesting slower, denser wind material along the line of sight and a wind structure focused toward the black hole. 

Furthermore, intensity-sorted spectroscopy of the dip phase indicates potential spectral variability on timescales of only a few seconds. Even with the limited statistical significance, the trends suggest that the wind ionization responds rapidly to changes in source flux, with higher ionized Fe absorptions during brighter intervals. This behavior implies interaction between the stellar wind and the X-ray source on few-second timescales and supports a picture in which a focused wind component coexists with clump-like structures of dense, less ionized material. The clumpy wind signatures identified in this study provide important clues to the driving mechanism of the stellar wind. Although, the intensity-sorted spectroscopy did not detect Fe\,K$\alpha$ emission variability at a significance exceeding the 1$\sigma$ level, the results motivate further dip observations in the hard state, as well as complementary soft-state observations, which should provide additional constraints on the origin of the Fe\,K$\alpha$ emission. 

Our results provide compact constraints for wind--accretion models in HMXBs: the data favor a structured outflow composed of a slower, dense focused component plus inhomogeneous clumps with distinct ionization and velocities. Smooth-wind prescriptions struggle to reproduce the observed orbital modulation and stochastic variability; instead, multi-dimensional radiation--hydrodynamics (and ultimately MHD) models with X-ray photoionization feedback, shielding/shadowing, and time-dependent clumping are required. Future work should pursue state- and phase-resolved XRISM/Resolve campaigns beyond the hard state--targeting intermediate and soft states--to test how the ionizing SED regulates the wind (e.g., Fe\,\textsc{xxv}/Fe\,\textsc{xxvi} absorption and Fe\,K$\alpha$ response). Our measurements provide benchmarks for complex (magneto)hydrodynamics simulations and for population-synthesis studies linking wind clumping and mass loss to compact-object masses and merger pathways.

The results of this paper place new constraints on the dynamics of wind-fed accretion in Cyg~X-1 and highlight the capability of XRISM/Resolve to reveal complex plasma conditions in high-mass X-ray binaries. Future XRISM campaigns, particularly those coordinated with multiwavelength observations, will further clarify the coupling between massive-star winds and black hole accretion.

\begin{ack}
This observation was made possible through the sustained dedication and invaluable contributions of the XRISM team, including scientists, engineers in detector and spacecraft systems, and the operations team. 
We are deeply grateful for their expertise, perseverance, and longstanding efforts in bringing this mission to fruition.
\end{ack}

\section*{Funding}
Part of this work was supported by JSPS KAKENHI grant Nos.\ JP24K00672 (S.Y.), 	23K22543 (S.Y.), JP21K13958 (M.M.), 23K25882 and 23H04895 (T.M.), 22H01269 (T.K.), 20H05857, Yamada Science Foundation (M.M.), and JSPS Core-to-Core Program (grant number:JPJSCCA20220002).
Part of this work was performed under the auspices of the U.S. Department of Energy by Lawrence Livermore National Laboratory under Contract DE-AC52-07NA27344. 

\section*{Data availability} 
The XRISM data underlying this article are available in heasarc data repository (NASA/GSFC) or DARTS (JAXA/ISAS). All the data reduction tools are available as a ftools package. 

\appendix

\section{Selection of the Xtend Source-Extraction Region}
Bright point sources observed with Xtend can suffer from event pile-up and grade migration, as well as detector effects (e.g., flickering pixels) and, at larger radii, contamination from dust-scattering halos (e.g., \cite{Xiang2011}). If unmitigated, these effects bias the soft X-ray spectrum and the inferred continuum/absorption parameters. We therefore quantified pile-up using the image-based method of \citet{Yamada2012-qx}, which estimates the pile-up fraction from the per-pixel count rate and grade distribution.

\begin{figure}[h]
  \begin{center}
    \includegraphics[width=\linewidth]{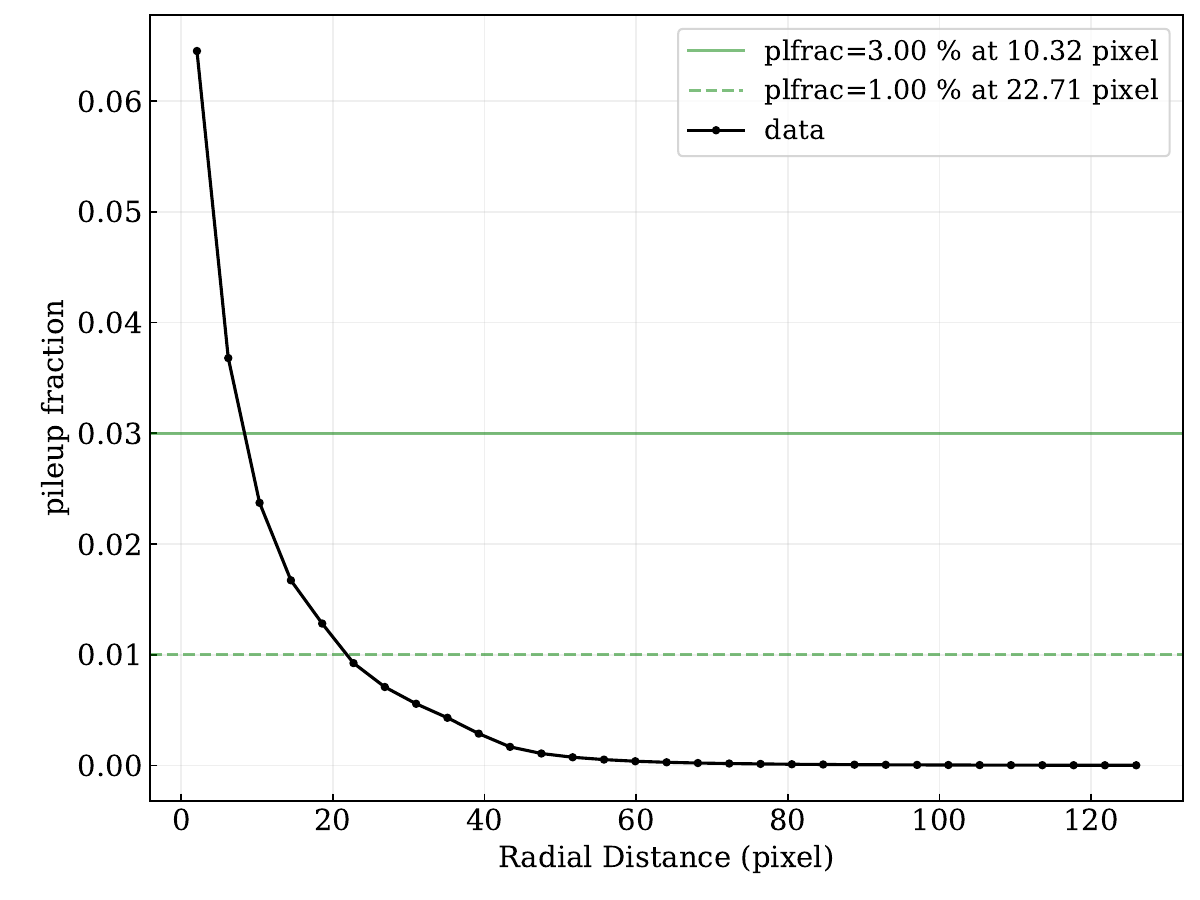}
  \end{center}
  \caption{Radial profile of the Xtend pile-up fraction. The fraction is $\sim$3\% at $r=10$~pixels and falls to $\sim$1\% by $r\approx22$~pixels. We therefore excise the central $r<10$~pixels and extract source events from an annulus (10--50~pixels).
{Alt text: This figure shows the pile-up fraction on the y-axis as a function of radial distance in pixels on the x-axis.}
  }
  \label{fig:appendix_pileup}
\end{figure}

Figure~\ref{fig:appendix_pileup} shows the radial profile of the pile-up fraction for the cleaned events used in this work. The fraction is $\sim$3\% within the central $r=10$~pixels and drops below $\sim$1\% beyond $r\approx22$~pixels. Balancing residual pile-up against signal-to-noise and encircled energy, we adopted an annular extraction region with inner radius $r_{\rm in}=10$~pixels and outer radius $r_{\rm out}=50$~pixels for the Xtend spectra (Section~\ref{sec:Observation and Data Reduction}). ARFs were generated for this geometry, accounting for the energy-dependent encircled-energy fraction; vignetting is negligible for our on-axis pointing. We repeated the extraction with $r_{\rm in}=\{8,12\}$~pixels and $r_{\rm out}=\{40,60\}$~pixels: after applying the corresponding ARFs, the spectra agree within statistical uncertainties over 0.4--10~keV, and best-fit continuum/line parameters remain consistent within errors. The non-X-ray background (NXB) is $<1$\% of the source signal across the band and was therefore not subtracted (Section~\ref{sec:Observation and Data Reduction}).

\section{Incident SED for the Photoionized Absorption Models \Cloudyabs\ and \pionabs}\label{Appendix:sec:SED}
The spectral energy distribution (SED) illuminating the absorber was constructed from joint XRISM/Xtend and NuSTAR spectra fitted with \texttt{tbabs}$\times$(\texttt{diskbb+nthComp}); see Section~\ref{subsec:Modeling of Fe Emission and Absorption Lines} for the best-fit parameters. We then removed the interstellar attenuation (\texttt{tbabs}) to obtain the intrinsic continuum seen by the wind. The \texttt{diskbb} and \texttt{nthComp} components were extrapolated over a broad energy range appropriate for photoionization (UV/soft X-rays to hard X-rays), using the analytic forms of the fitted models; the relative normalizations were fixed to the joint best fit. 

\begin{figure}[h]
  \begin{center}
    \includegraphics[width=\linewidth]{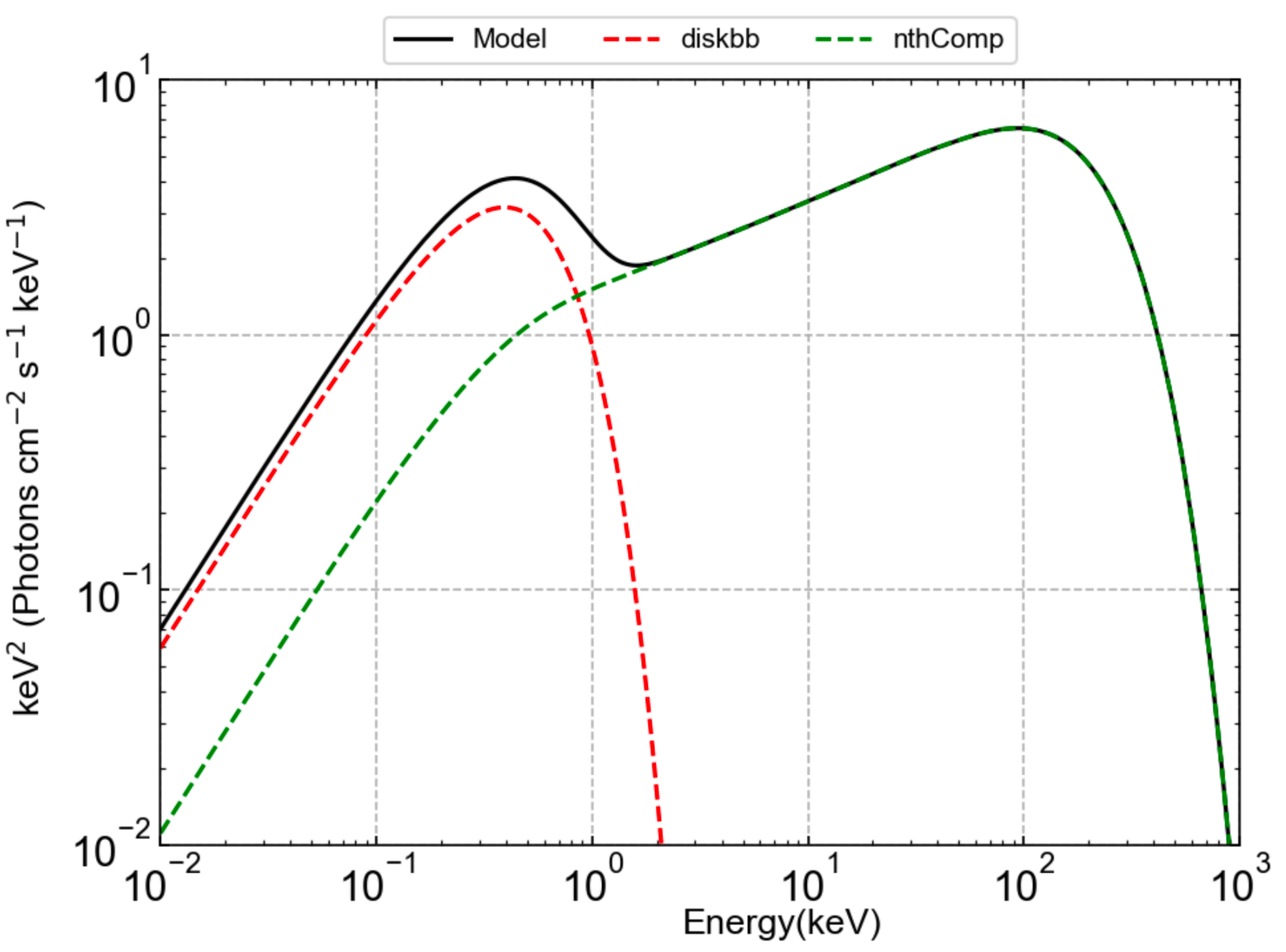}
  \end{center}
  \caption{Incident continuum used in the photoionization calculations. The \texttt{diskbb} and \texttt{nthComp} components are shown by red and green dashed curves, respectively; the black solid curve is their sum (interstellar absorption removed). This SED was supplied to \Cloudyabs\ and \pionabs\ for computing the ionization balance and line opacities.
{Alt text: This figure shows X-ray spectral energy distribution with energy on the x-axis and photon flux (keV$^{2}$ photons cm$^{-2}$ s$^{-1}$ keV$^{-1}$) on the y-axis.}
  }
  \label{fig:appendix_cloudy_sed}
\end{figure}

Figure~\ref{fig:appendix_cloudy_sed} shows the resulting SED: the multi-temperature disk (\textit{red dashed}), the Comptonized continuum (\textit{green dashed}), and their sum (\textit{black}). This SED was provided to \Cloudyabs\ (generated with \textsc{Cloudy}; e.g., \cite{Chatzikos2023-ym,Gunasekera2023-vr}) and to the \texttt{pion}-based absorber (\pionabs) in \textsc{SPEX}. Each code used its default abundance set and atomic data; differences in element tables and radiative transfer lead to small systematic offsets in $(\log\xi,N({\mathrm{H}}))$ (Section~\ref{subsec:Modeling of Fe Emission and Absorption Lines}), but the predicted line profiles in the Fe\,K band are consistent. For fitting, we computed model grids spanning the parameter ranges relevant to our spectra and interpolated within those grids. The microturbulent velocity was treated as a free parameter. The same SED was used for all fitting used in this paper. 

\bibliographystyle{aasjournal}
\bibliography{Reference, Cyg-X-1_cvt,XRISM_cvt,ASTRO-H_cvt,Calorimeter_cvt,Plasma-Physics_cvt,Calorimeter_cvt_spie,Cyg-X-3_cvt,Cen-X-3_cvt}

@article{Holzer1997-cr,
 author = {H{\"o}lzer, G and Fritsch, M and Deutsch, M and H{\"a}rtwig, J and
F{\"o}rster, E},
 doi = {10.1103/PhysRevA.56.4554},
 journal = {Phys. Rev. A},
 month = {December},
 pages = {4554--4568},
 publisher = {American Physical Society},
 title = {$K\ensuremath\alpha_1,2$ and $K\ensuremath\beta_1,3$ x-ray
emission lines of the $3d$ transition metals},
 url = {https://link.aps.org/doi/10.1103/PhysRevA.56.4554},
 volume = {56},
 year = {1997}
}

@article{Rudolph2013-vu,
 archiveprefix = {arXiv},
 author = {Rudolph, J K and Bernitt, S and Epp, S W and Steinbr{\"u}gge, R
and Beilmann, C and Brown, G V and Eberle, S and Graf, A and
Harman, Z and Hell, N and Leutenegger, M and M{\"u}ller, A and
Schlage, K and Wille, H-C and Yavaş, H and Ullrich, J and
Crespo L{\'o}pez-Urrutia, J R},
 doi = {10.1103/PhysRevLett.111.103002},
 eprint = {1306.4348},
 issn = {0031-9007,1079-7114},
 journal = {Phys. Rev. Lett.},
 month = {September},
 number = {10},
 pages = {103002},
 primaryclass = {physics.atom-ph},
 title = {{X}-Ray Resonant Photoexcitation: Linewidths and Energies of
K\ensuremath\alpha Transitions in Highly Charged Fe Ions},
 url = {http://dx.doi.org/10.1103/PhysRevLett.111.103002},
 volume = {111},
 year = {2013}
}

@article{Mochizuki2024-yz,
 author = {Mochizuki, Yuto and Tsujimoto, Masahiro and Kelley, Richard L and
Vander Meulen, Bert and Enoto, Teruaki and Nagai, Yutaro and
Done, Chris and Pradhan, Pragati and Hell, Natalie and
Pottschmidt, Katja and Ebisawa, Ken and Behar, Ehud},
 doi = {10.3847/2041-8213/ad946d},
 issn = {2041-8205,2041-8213},
 journal = {Astrophys. J. Lett.},
 language = {en},
 month = {10~December},
 number = {1},
 pages = {L21},
 publisher = {American Astronomical Society},
 title = {Detection of the orbital modulation of Fe Kα fluorescence
emission in Centaurus {X}-3 using the high-resolution
spectrometer resolve on board {XRISM}},
 url = {http://dx.doi.org/10.3847/2041-8213/ad946d},
 volume = {977},
 year = {2024}
}

@article{Brocksopp1999-zn,
 archiveprefix = {arXiv},
 author = {Brocksopp, C and Tarasov, A E and Lyuty, V M and Roche, P},
 doi = {10.48550/arXiv.astro-ph/9812077},
 eprint = {astro-ph/9812077},
 issn = {0004-6361,1432-0746},
 journal = {Astron. Astrophys.},
 month = {March},
 pages = {861--864},
 primaryclass = {astro-ph},
 title = {An improved orbital ephemeris for Cygnus {X}-1},
 url = {http://dx.doi.org/10.48550/arXiv.astro-ph/9812077},
 volume = {343},
 year = {1999a}
}

@article{Draghis2024-oz,
 archiveprefix = {arXiv},
 author = {Draghis, Paul A and Miller, Jon M and Costantini, Elisa and
Gallo, Luigi C and Reynolds, Mark and Tomsick, John A and
Zoghbi, Abderahmen},
 doi = {10.3847/1538-4357/ad43ea},
 eprint = {2311.16225},
 issn = {0004-637X,1538-4357},
 journal = {Astrophys. J.},
 month = {July},
 number = {1},
 pages = {40},
 primaryclass = {astro-ph.HE},
 title = {Systematically Revisiting All {NuSTAR} Spins of Black Holes
in {X}-Ray Binaries},
 url = {http://dx.doi.org/10.3847/1538-4357/ad43ea},
 volume = {969},
 year = {2024}
}

@article{Friend1982-qb,
 author = {Friend, D B and Castor, J I},
 doi = {10.1086/160340},
 issn = {0004-637X,1538-4357},
 journal = {Astrophys. J.},
 month = {October},
 pages = {293},
 publisher = {IOP Publishing},
 title = {Radiation-driven winds in {X}-ray binaries},
 url = {https://ui.adsabs.harvard.edu/abs/1982ApJ...261..293F/abstract},
 volume = {261},
 year = {1982}
}

@article{Gies1986-bz,
 author = {Gies, D R and Bolton, C T},
 doi = {10.1086/164172},
 issn = {0004-637X,1538-4357},
 journal = {Astrophys. J.},
 language = {en},
 month = {May},
 pages = {389},
 publisher = {American Astronomical Society},
 title = {The optical spectrum of {HDE} {226868=Cygnus} {X}-1. {III}. A
focused stellar wind model for he {II} lambda 4686 emission},
 url = {https://ui.adsabs.harvard.edu/abs/1986ApJ...304..389G/abstract},
 volume = {304},
 year = {1986}
}

@article{Grinberg2015-kl,
 author = {Grinberg, V and Leutenegger, M A and Hell, N and Pottschmidt, K
and B{\"o}ck, M and Garc{\'i}a, J A and Hanke, M and Nowak, M A and
Sundqvist, J O and Townsend, R H D and Wilms, J},
 doi = {10.1051/0004-6361/201425418},
 issn = {0004-6361,1432-0746},
 journal = {Astron. Astrophys.},
 month = {April},
 pages = {A117},
 publisher = {EDP Sciences},
 title = {Long term variability of Cygnus {X}-1: {VII}. Orbital variability
of the focussed wind in Cyg {X}-1/{HDE} 226868 system⋆},
 url = {https://www.aanda.org/articles/aa/full_html/2015/04/aa25418-14/aa25418-14.html},
 volume = {576},
 year = {2015}
}

@article{Hanke2009-rh,
 author = {Hanke, Manfred and Wilms, J{\"o}rn and Nowak, Michael A and
Pottschmidt, Katja and Schulz, Norbert S and Lee, Julia C},
 doi = {10.1088/0004-637x/690/1/330},
 issn = {0004-637X,1538-4357},
 journal = {Astrophys. J.},
 month = {1~January},
 number = {1},
 pages = {330--346},
 publisher = {IOP Publishing},
 title = {Chandrax-ray spectroscopy of the focused wind in the Cygnus x-1
system. {I}. The nondip spectrum in the low/hard state},
 url = {https://iopscience.iop.org/article/10.1088/0004-637X/690/1/330},
 volume = {690},
 year = {2009}
}

@article{Harer2023-gc,
 author = {H{\"a}rer, L K and Parker, M L and El Mellah, I and Grinberg, V and
Ballhausen, R and Igo, Z and Joyce, A and Wilms, J},
 doi = {10.1051/0004-6361/202346669},
 issn = {0004-6361,1432-0746},
 journal = {Astron. Astrophys.},
 month = {24~October},
 pages = {A72},
 publisher = {EDP Sciences},
 title = {Stellar-wind variability in Cygnus {X}-1 from high-resolution
excess variance spectroscopy with Chandra},
 url = {https://www.aanda.org/articles/aa/full_html/2023/12/aa46669-23/aa46669-23.html},
 volume = {680},
 year = {2023}
}

@article{Herrero1995-oh,
 author = {Herrero, A and Kudritzki, R P and Gabler, R and Vilchez, J M and
Gabler, A},
 issn = {0004-6361,1432-0746},
 journal = {Astron. Astrophys.},
 month = {May},
 pages = {556},
 title = {Fundamental parameters of galactic luminous {OB} stars. {II}. A
spectroscopic analysis of {HDE} 226868 and the mass of Cygnus
{X}-1},
 volume = {297},
 year = {1995}
}

@article{Hirsch2019-em,
 author = {Hirsch, Maria and Hell, Natalie and Grinberg, Victoria and
Ballhausen, Ralf and Nowak, Michael A and Pottschmidt, Katja and
Schulz, Norbert S and Dauser, Thomas and Hanke, Manfred and
Kallman, Timothy R and Brown, Gregory V and Wilms, J{\"o}rn},
 doi = {10.1051/0004-6361/201935074},
 issn = {0004-6361,1432-0746},
 journal = {Astron. Astrophys.},
 month = {June},
 pages = {A64},
 publisher = {EDP Sciences},
 title = {\textit{Chandra}X-ray spectroscopy of the focused wind in the
Cygnus {X}-1 system: {III}. Dipping in the low/hard state},
 url = {https://www.aanda.org/articles/aa/full_html/2019/06/aa35074-19/aa35074-19.html},
 volume = {626},
 year = {2019}
}

@article{Kitamoto1984-hc,
 author = {Kitamoto, S and Miyamoto, S and Tanaka, Yasuyuki T and Ohashi, T
and Kondo, Y and Tawara, Y and Nakagawa, M},
 issn = {0004-6264,0004-6264},
 journal = {\pasj},
 pages = {731--740},
 title = {Transient dips of Cygnus {X}-1 observed from Tenma},
 volume = {36},
 year = {1984}
}

@article{Makishima2008-wn,
 author = {Makishima, Kazuo and Takahashi, Hiromitsu and Yamada, Shin'ya and
Done, Chris and Kubota, Aya and Dotani, Tadayasu and Ebisawa, Ken
and Itoh, Takeshi and Kitamoto, Shunji and Negoro, Hitoshi and
Ueda, Yoshihiro and Yamaoka, Kazutaka},
 doi = {10.1093/pasj/60.3.585},
 issn = {0004-6264,2053-051X},
 journal = {Publ. Astron. Soc. Jpn. Nihon Tenmon Gakkai},
 language = {en},
 month = {25~June},
 number = {3},
 pages = {585--604},
 publisher = {Oxford University Press (OUP)},
 title = {Suzaku results on Cygnus {X}-1 in the low/hard state},
 url = {https://academic.oup.com/pasj/article/60/3/585/2332822},
 volume = {60},
 year = {2008}
}

@article{Miller-Jones2021-nt,
 author = {Miller-Jones, James C A and Bahramian, Arash and Orosz, Jerome A
and Mandel, Ilya and Gou, Lijun and Maccarone, Thomas J and
Neijssel, Coenraad J and Zhao, Xueshan and Zi{\'o}łkowski, Janusz and
Reid, Mark J and Uttley, Phil and Zheng, Xueying and Byun,
Do-Young and Dodson, Richard and Grinberg, Victoria and Jung,
Taehyun and Kim, Jeong-Sook and Marcote, Benito and Markoff, Sera
and Rioja, Mar{\'i}a J and Rushton, Anthony P and Russell, David M
and Sivakoff, Gregory R and Tetarenko, Alexandra J and Tudose,
Valeriu and Wilms, Joern},
 doi = {10.1126/science.abb3363},
 issn = {0036-8075,1095-9203},
 journal = {Science},
 language = {en},
 month = {5~March},
 number = {6533},
 pages = {1046--1049},
 pmid = {33602863},
 publisher = {American Association for the Advancement of Science (AAAS)},
 title = {Cygnus {X}-1 contains a 21-solar mass black hole-Implications for
massive star winds},
 url = {https://www.science.org/doi/10.1126/science.abb3363},
 volume = {371},
 year = {2021}
}

@article{Miller2005-sw,
 author = {Miller, J M and Wojdowski, P and Schulz, N S and Marshall, H L
and Fabian, A C and Remillard, R A and Wijnands, R and Lewin, W H
G},
 doi = {10.1086/426701},
 issn = {0004-637X,1538-4357},
 journal = {Astrophys. J.},
 language = {en},
 month = {10~February},
 number = {1},
 pages = {398--404},
 publisher = {American Astronomical Society},
 title = {Revealing the focused companion wind in Cygnus {X‐1}
{withChandra}},
 url = {https://iopscience.iop.org/article/10.1086/426701},
 volume = {620},
 year = {2005}
}

@article{Miskovicova2016-zi,
 author = {Mi{\v{s}}kovi{\v{c}}ov{\'a}, Ivica and Hell, Natalie and Hanke, Manfred and
Nowak, Michael A and Pottschmidt, Katja and Schulz, Norbert S and
Grinberg, Victoria and Duro, Refiz and Madej, Oliwia K and
Lohfink, Anne M and Rodriguez, J{\'e}r{\^o}me and Bel, Marion Cadolle and
Bodaghee, Arash and Tomsick, John A and Lee, Julia C and Brown,
Gregory V and Wilms, J{\"o}rn},
 doi = {10.1051/0004-6361/201322490},
 issn = {0004-6361,1432-0746},
 journal = {Astron. Astrophys.},
 month = {June},
 pages = {A114},
 publisher = {EDP Sciences},
 title = {{ChandraX}-ray spectroscopy of focused wind in the Cygnus {X}-1
system: {II}. The non-dip spectrum in the low/hard state -
modulations with orbital phase},
 url = {https://www.aanda.org/articles/aa/abs/2016/06/aa22490-13/aa22490-13.html},
 volume = {590},
 year = {2016}
}

@article{Orosz2011-ez,
 archiveprefix = {arXiv},
 author = {Orosz, Jerome A and McClintock, Jeffrey E and Aufdenberg,
Jason P and Remillard, Ronald A and Reid, Mark J and Narayan,
Ramesh and Gou, Lijun},
 doi = {10.1088/0004-637X/742/2/84},
 eprint = {1106.3689},
 issn = {0004-637X,1538-4357},
 journal = {Astrophys. J.},
 month = {December},
 number = {2},
 pages = {84},
 primaryclass = {astro-ph.HE},
 title = {The Mass of the Black Hole in Cygnus {X}-1},
 url = {http://dx.doi.org/10.1088/0004-637X/742/2/84},
 volume = {742},
 year = {2011}
}

@article{Ramachandran2025-jn,
 archiveprefix = {arXiv},
 author = {Ramachandran, V and Sander, A A C and Oskinova, L M and
Sch{\"o}sser, E C and Pauli, D and Hamann, W-R and Mahy, L and
Bernini-Peron, M and Brigitte, M and Kub{\'a}tov{\'a}, B},
 doi = {10.1051/0004-6361/202554184},
 eprint = {2504.05885},
 issn = {0004-6361,1432-0746},
 journal = {Astron. Astrophys.},
 month = {May},
 pages = {A37},
 primaryclass = {astro-ph.SR},
 title = {Comprehensive {UV} and optical spectral analysis of Cygnus
{X}-1: Stellar and wind parameters, abundances, and
evolutionary implications},
 url = {http://dx.doi.org/10.1051/0004-6361/202554184},
 volume = {698},
 year = {2025}
}

@article{Tomsick2018-qk,
 archiveprefix = {arXiv},
 author = {Tomsick, John A and Parker, Michael L and Garc{\'i}a, Javier A
and Yamaoka, Kazutaka and Barret, Didier and Chiu, Jeng-Lun
and Clavel, Ma{\"i}ca and Fabian, Andrew and F{\"u}rst, Felix and
Gandhi, Poshak and Grinberg, Victoria and Miller, Jon M and
Pottschmidt, Katja and Walton, Dominic J},
 doi = {10.3847/1538-4357/aaaab1},
 eprint = {1801.07267},
 issn = {0004-637X,1538-4357},
 journal = {Astrophys. J.},
 month = {March},
 number = {1},
 pages = {3},
 primaryclass = {astro-ph.HE},
 title = {Alternative Explanations for Extreme Supersolar Iron
Abundances Inferred from the Energy Spectrum of Cygnus {X}-1},
 url = {http://dx.doi.org/10.3847/1538-4357/aaaab1},
 volume = {855},
 year = {2018}
}

@article{Yamada2012-qx,
 author = {Yamada, Shin'ya and Uchiyama, Hideki and Dotani, Tadayasu and
Tsujimoto, Masahiro and Katsuda, Satoru and Makishima, Kazuo and
Takahashi, Hiromitsu and Noda, Hirofumi and Torii, Shunsuke and
Sakurai, Soki and Enoto, Teruaki and Yuasa, Takayuki and Koyama,
Shu and Bamba, Aya},
 doi = {10.1093/pasj/64.3.53},
 issn = {0004-6264,2053-051X},
 journal = {Publ. Astron. Soc. Jpn. Nihon Tenmon Gakkai},
 language = {en},
 month = {25~June},
 number = {3},
 pages = {53},
 publisher = {Oxford University Press (OUP)},
 title = {Data-oriented diagnostics of pileup effects on the suzaku {XIS}},
 url = {https://academic.oup.com/pasj/article/64/3/53/1595049},
 volume = {64},
 year = {2012}
}

@article{Yamada2013-wg,
 author = {Yamada, Shin'ya and Makishima, Kazuo and Done, Chris and Torii,
Shunsuke and Noda, Hirofumi and Sakurai, Soki},
 doi = {10.1093/pasj/65.4.80},
 issn = {0004-6264,2053-051X},
 journal = {Publ. Astron. Soc. Jpn. Nihon Tenmon Gakkai},
 language = {en},
 month = {25~August},
 number = {4},
 pages = {80},
 publisher = {Oxford University Press (OUP)},
 title = {Evidence for a cool disk and inhomogeneous coronae from wide-band
temporal spectroscopy of Cygnus {X}-1 with suzaku},
 url = {https://academic.oup.com/pasj/article/65/4/80/2898351},
 volume = {65},
 year = {2013}
}

@article{Zdziarski2024-mi,
 author = {Zdziarski, Andrzej A and Banerjee, Srimanta and Chand, Swadesh
and Dewangan, Gulab and Misra, Ranjeev and Szanecki, Michał and
Niedźwiecki, Andrzej},
 doi = {10.3847/1538-4357/ad1b60},
 issn = {0004-637X,1538-4357},
 journal = {Astrophys. J.},
 month = {1~February},
 number = {2},
 pages = {101},
 publisher = {American Astronomical Society},
 title = {Black hole spin measurements in {LMC} {X}-1 and Cyg {X}-1 are
highly model dependent},
 url = {http://dx.doi.org/10.3847/1538-4357/ad1b60},
 volume = {962},
 year = {2024}
}

@article{XRISM-Collaboration2024-vs,
 author = {{XRISM Collaboration} and Audard, Marc and Awaki, Hisamitsu and
Ballhausen, Ralf and Bamba, Aya and Behar, Ehud and
Boissay-Malaquin, Rozenn and Brenneman, Laura and Brown, Gregory
V and Corrales, Lia and Costantini, Elisa and Cumbee, Renata and
D{\'i}az Trigo, Mar{\'i}a and Done, Chris and Dotani, Tadayasu and
Ebisawa, Ken and Eckart, Megan E and Eckert, Dominique and
Eguchi, Satoshi and Enoto, Teruaki and Ezoe, Yuichiro and Foster,
Adam and Fujimoto, Ryuichi and Fujita, Yutaka and Fukazawa,
Yasushi and Fukushima, Kotaro and Furuzawa, Akihiro and Gallo,
Luigi and Garc{\'i}a, Javier A and Gu, Liyi and Guainazzi, Matteo and
Hagino, Kouichi and Hamaguchi, Kenji and Hatsukade, Isamu and
Hayashi, Katsuhiro and Hayashi, Takayuki and Hell, Natalie and
Hodges-Kluck, Edmund and Hornschemeier, Ann and Ichinohe, Yuto
and Ishida, Manabu and Ishikawa, Kumi and Ishisaki, Yoshitaka and
Kaastra, Jelle and Kallman, Timothy and Kara, Erin and Katsuda,
Satoru and Kanemaru, Yoshiaki and Kelley, Richard and Kilbourne,
Caroline and Kitamoto, Shunji and Kobayashi, Shogo and Kohmura,
Takayoshi and Kubota, Aya and Leutenegger, Maurice and
Loewenstein, Michael and Maeda, Yoshitomo and Markevitch, Maxim
and Matsumoto, Hironori and Matsushita, Kyoko and McCammon, Dan
and McNamara, Brian and Mernier, Fran{\c c}ois and Miller, Eric D and
Miller, Jon M and Mitsuishi, Ikuyuki and Mizumoto, Misaki and
Mizuno, Tsunefumi and Mori, Koji and Mukai, Koji and Murakami,
Hiroshi and Mushotzky, Richard and Nakajima, Hiroshi and
Nakazawa, Kazuhiro and Ness, Jan-Uwe and Nobukawa, Kumiko and
Nobukawa, Masayoshi and Noda, Hirofumi and Odaka, Hirokazu and
Ogawa, Shoji and Ogorzalek, Anna and Okajima, Takashi and Ota,
Naomi and Paltani, Stephane and Petre, Robert and Plucinsky, Paul
and Porter, Frederick S and Pottschmidt, Katja and Sato, Kosuke
and Sato, Toshiki and Sawada, Makoto and Seta, Hiromi and
Shidatsu, Megumi and Simionescu, Aurora and Smith, Randall and
Suzuki, Hiromasa and Szymkowiak, Andrew and Takahashi, Hiromitsu
and Takeo, Mai and Tamagawa, Toru and Tamura, Keisuke and Tanaka,
Takaaki and Tanimoto, Atsushi and Tashiro, Makoto and Terada,
Yukikatsu and Terashima, Yuichi and Tsuboi, Yohko and Tsujimoto,
Masahiro and Tsunemi, Hiroshi and Tsuru, Takeshi and Uchida,
Hiroyuki and Uchida, Nagomi and Uchida, Yuusuke and Uchiyama,
Hideki and Ueda, Yoshihiro and Uno, Shinichiro and Vink, Jacco
and Watanabe, Shin and Williams, Brian J and Yamada, Satoshi and
Yamada, Shinya and Yamaguchi, Hiroya and Yamaoka, Kazutaka and
Yamasaki, Noriko and Yamauchi, Makoto and Yamauchi, Shigeo and
Yaqoob, Tahir and Yoneyama, Tomokage and Yoshida, Tessei and
Yukita, Mihoko and Zhuravleva, Irina and Tomaru, Ryota and
Hayashi, Tasuku and Hakamata, Tomohiro and Miura, Daiki and
Koljonen, Karri and McCollough, Mike},
 doi = {10.3847/2041-8213/ad8ed0},
 issn = {2041-8205,2041-8213},
 journal = {Astrophys. J. Lett.},
 month = {20~December},
 number = {2},
 pages = {L34},
 publisher = {American Astronomical Society},
 title = {The {XRISM}/Resolve view of the Fe {K} region of Cyg {X}-3},
 url = {https://iopscience.iop.org/article/10.3847/2041-8213/ad8ed0/pdf},
 volume = {977},
 year = {2024}
}

@article{Castor1975-wm,
 author = {Castor, J I and Abbott, D C and Klein, R I},
 doi = {10.1086/153315},
 issn = {0004-637X,1538-4357},
 journal = {Astrophys. J.},
 month = {January},
 pages = {157},
 publisher = {IOP Publishing},
 title = {Radiation-driven winds in Of stars},
 url = {http://adsabs.harvard.edu/doi/10.1086/153315},
 volume = {195},
 year = {1975}
}

@article{Chatzikos2023-ym,
 archiveprefix = {arXiv},
 author = {Chatzikos, M and Bianchi, S and Camilloni, F and Chakraborty,
P and Gunasekera, C M and Guzm{\'a}n, F and Milby, J S and
Sarkar, A and Shaw, G and van Hoof, P A M and Ferland, G J},
 doi = {10.22201/ia.01851101p.2023.59.02.12},
 eprint = {2308.06396},
 issn = {0185-1101},
 journal = {Rev. Mex. Astron. Astrofis.},
 month = {October},
 pages = {327--343},
 primaryclass = {astro-ph.GA},
 title = {The 2023 Release of Cloudy},
 url = {http://dx.doi.org/10.22201/ia.01851101p.2023.59.02.12},
 volume = {59},
 year = {2023}
}

@article{Gunasekera2023-vr,
 archiveprefix = {arXiv},
 author = {Gunasekera, Chamani M and van Hoof, Peter A M and Chatzikos,
Marios and Ferland, Gary J},
 doi = {10.3847/2515-5172/ad0e75},
 eprint = {2311.10163},
 journal = {Research Notes of the American Astronomical Society},
 month = {November},
 number = {11},
 pages = {246},
 primaryclass = {astro-ph.GA},
 title = {The 23.01 Release of Cloudy},
 url = {http://dx.doi.org/10.3847/2515-5172/ad0e75},
 volume = {7},
 year = {2023}
}

@article{Ueda2004-tv,
 author = {Ueda, Y and Murakami, H and Yamaoka, K and Dotani, T and Ebisawa,
K},
 doi = {10.1086/420973},
 journal = {The Astrophysical Journal},
 month = {July},
 number = {1},
 pages = {325},
 title = {Chandra High-Resolution Spectroscopy of the Absorption-Line
Features in the Low-Mass {X}-Ray Binary {GX} {13+1}},
 url = {https://dx.doi.org/10.1086/420973},
 volume = {609},
 year = {2004}
}

@ARTICLE{Oda1971,
       author = {{Oda}, M. and {Gorenstein}, P. and {Gursky}, H. and {Kellogg}, E. and {Schreier}, E. and {Tananbaum}, H. and {Giacconi}, R.},
        title = "{X-Ray Pulsations from Cygnus X-1 Observed from UHURU}",
      journal = {\apjl},
         year = 1971,
        month = may,
       volume = {166},
        pages = {L1},
          doi = {10.1086/180726},
       adsurl = {https://ui.adsabs.harvard.edu/abs/1971ApJ...166L...1O}
}

@article{Miura2025,
    author = {Miura, Daiki and Yamaguchi, Hiroya and Ballhausen, Ralf and Kallman, Timothy and Enoto, Teruaki and Yamada, Shinya and Hakamata, Tomohiro and Tomaru, Ryota and Odaka, Hirokazu and Hell, Natalie and Nakajima, Hiroshi and Watanabe, Shin and Hayashi, Tasuku and Kitamoto, Shunji and Yamaoka, Kazutaka and Miller, Jon M and Okabe, Keigo and Maruzuka, Itsuki and Koljonen, Karri and McCollough, Mike},
    title = {XRISM spectroscopy on orbital modulation of Fe Lyα lines in Cygnus X-3},
    journal = {\pasj},
    volume = {77},
    number = {Supplement_1},
    pages = {S86-S95},
    year = {2025},
    month = {06},
    issn = {2053-051X},
    doi = {10.1093/pasj/psaf057},
    url = {https://doi.org/10.1093/pasj/psaf057},
    eprint = {https://academic.oup.com/pasj/article-pdf/77/Supplement_1/S86/63496808/psaf057.pdf},
}

@ARTICLE{Yamada2025,
       author = {{Yamada}, Shinya and {Hell}, Natalie and {Costantini}, Elisa and {Adegoke}, Oluwashina and {Brumback}, McKinley and {Draghis}, Paul and {Ebisawa}, Ken and {Garcia}, Javier A. and {Hodges-Kluck}, Edmund and {Kitamoto}, Shunji and {Kobayashi}, Shogo and {Kohmura}, Takayoshi and {Kubota}, Aya and {Miller}, Jon M. and {Mizumoto}, Misaki and {Mizuno}, Tsunefumi and {Ninoyu}, Kaito and {Takahashi}, Hiromitsu and {Uchida}, Yuusuke and {Yamaoka}, Kazutaka and {Zhang}, Sixuan},
        title = "{XRISM high-resolution X-ray spectroscopy of Cygnus X-1: Highly ionized iron absorption structures}",
      journal = {\pasj},
         year = 2025,
        month = dec,
       volume = {77},
       number = {6},
        pages = {1210-1223},
          doi = {10.1093/pasj/psaf104},
       adsurl = {https://ui.adsabs.harvard.edu/abs/2025PASJ...77.1210Y}
}

@ARTICLE{Miller2002,
       author = {{Miller}, J.~M. and {Fabian}, A.~C. and {Wijnands}, R. and {Remillard}, R.~A. and {Wojdowski}, P. and {Schulz}, N.~S. and {Di Matteo}, T. and {Marshall}, H.~L. and {Canizares}, C.~R. and {Pooley}, D. and {Lewin}, W.~H.~G.},
        title = "{Resolving the Composite Fe K{\ensuremath{\alpha}} Emission Line in the Galactic Black Hole Cygnus X-1 with Chandra}",
      journal = {\apj},
         year = 2002,
        month = oct,
       volume = {578},
       number = {1},
        pages = {348-356},
          doi = {10.1086/342466},
archivePrefix = {arXiv},
       eprint = {astro-ph/0202083},
 primaryClass = {astro-ph},
       adsurl = {https://ui.adsabs.harvard.edu/abs/2002ApJ...578..348M}
}

@ARTICLE{Tananbaum1972,
       author = {{Tananbaum}, H. and {Gursky}, H. and {Kellogg}, E. and {Giacconi}, R. and {Jones}, C.},
        title = "{Observation of a Correlated X-Ray Transition in Cygnus X-1}",
      journal = {\apjl},
         year = 1972,
        month = oct,
       volume = {177},
        pages = {L5},
          doi = {10.1086/181042},
       adsurl = {https://ui.adsabs.harvard.edu/abs/1972ApJ...177L...5T}
}

@ARTICLE{Nowak2011,
       author = {{Nowak}, Michael A. and {Hanke}, Manfred and {Trowbridge}, Sarah N. and {Markoff}, Sera B. and {Wilms}, J{\"o}rn and {Pottschmidt}, Katja and {Coppi}, Paolo and {Maitra}, Dipankar and {Davis}, John E. and {Tramper}, Frank},
        title = "{Corona, Jet, and Relativistic Line Models for Suzaku/RXTE/Chandra-HETG Observations of the Cygnus X-1 Hard State}",
      journal = {\apj},
         year = 2011,
        month = feb,
       volume = {728},
       number = {1},
          eid = {13},
        pages = {13},
          doi = {https://doi.org/10.1088/0004-637X/728/1/13},
archivePrefix = {arXiv},
       eprint = {https://ui.adsabs.harvard.edu/abs/2011ApJ...728...13N},
 primaryClass = {astro-ph.HE},
       adsurl = {https://ui.adsabs.harvard.edu/abs/2011ApJ...728...13N}
}

@article{Fabian2012,
	adsurl = {https://ui.adsabs.harvard.edu/abs/2012MNRAS.424..217F},
	archiveprefix = {arXiv},
	author = {{Fabian}, A.~C. and {Wilkins}, D.~R. and {Miller}, J.~M. and {Reis}, R.~C. and {Reynolds}, C.~S. and {Cackett}, E.~M. and {Nowak}, M.~A. and {Pooley}, G.~G. and {Pottschmidt}, K. and {Sanders}, J.~S. and {Ross}, R.~R. and {Wilms}, J.},
	doi = {https://doi.org/10.1111/j.1365-2966.2012.21185.x},
	eprint = {1204.5854},
	journal = {\mnras},
	month = jul,
	number = {1},
	pages = {217-223},
	primaryclass = {astro-ph.HE},
	title = {{On the determination of the spin of the black hole in Cyg X-1 from X-ray reflection spectra}},
	volume = {424},
	year = 2012}

@article{Basak2017,
       author = {{Basak}, Rupal and {Zdziarski}, Andrzej A. and {Parker}, Michael and {Islam}, Nazma},
        title = "{Analysis of NuSTAR and Suzaku observations of Cyg X-1 in the hard state: evidence for a truncated disc geometry}",
      journal = {\mnras},
         year = 2017,
        month = dec,
       volume = {472},
       number = {4},
        pages = {4220-4232},
          doi = {https://dx.doi.org/10.1093/mnras/stx2283},
archivePrefix = {arXiv},
       eprint = {https://academic.oup.com/mnras/article/472/4/4220/4104653},
 primaryClass = {astro-ph.HE},
       adsurl = {https://ui.adsabs.harvard.edu/abs/2017MNRAS.472.4220B}
}

@ARTICLE{Done2007,
      author = {{Done}, Chris and {Gierli{\'n}ski}, Marek and {Kubota}, Aya},
        title = "{Modelling the behaviour of accretion flows in X-ray binaries. Everything you always wanted to know about accretion but were afraid to ask}",
      journal = {\aapr},
         year = 2007,
        month = dec,
       volume = {15},
       number = {1},
        pages = {1-66},
          doi = {https://doi.org/10.1007/s00159-007-0006-1},
archivePrefix = {arXiv},
       eprint = {https://ui.adsabs.harvard.edu/abs/2007A&ARv..15....1D},
 primaryClass = {astro-ph},
       adsurl = {https://ui.adsabs.harvard.edu/abs/2007A&ARv..15....1D}
}

@ARTICLE{Walborn1973,
       author = {{Walborn}, N.~R.},
        title = "{The space distribution of the O stars in the solar neighborhood.}",
      journal = {\aj},
         year = 1973,
        month = dec,
       volume = {78},
        pages = {1067-1083},
          doi = {10.1086/111509},
       adsurl = {https://ui.adsabs.harvard.edu/abs/1973AJ.....78.1067W}
}

@ARTICLE{Owocki1988,
       author = {{Owocki}, Stanley P. and {Castor}, John I. and {Rybicki}, George B.},
        title = "{Time-dependent Models of Radiatively Driven Stellar Winds. I. Nonlinear Evolution of Instabilities for a Pure Absorption Model}",
      journal = {\apj},
         year = 1988,
        month = dec,
       volume = {335},
        pages = {914},
          doi = {10.1086/166977},
       adsurl = {https://ui.adsabs.harvard.edu/abs/1988ApJ...335..914O}
}

@ARTICLE{Sundqvist2018,
       author = {{Sundqvist}, J.~O. and {Owocki}, S.~P. and {Puls}, J.},
        title = "{2D wind clumping in hot, massive stars from hydrodynamical line-driven instability simulations using a pseudo-planar approach}",
      journal = {\aap},
         year = 2018,
        month = mar,
       volume = {611},
          eid = {A17},
        pages = {A17},
          doi = {10.1051/0004-6361/201731718},
archivePrefix = {arXiv},
       eprint = {1710.07780},
 primaryClass = {astro-ph.SR},
       adsurl = {https://ui.adsabs.harvard.edu/abs/2018A&A...611A..17S}
}

@ARTICLE{Oskinova2012,
       author = {{Oskinova}, L.~M. and {Feldmeier}, A. and {Kretschmar}, P.},
        title = "{Clumped stellar winds in supergiant high-mass X-ray binaries: X-ray variability and photoionization}",
      journal = {\mnras},
         year = 2012,
        month = apr,
       volume = {421},
       number = {4},
        pages = {2820-2831},
          doi = {10.1111/j.1365-2966.2012.20507.x},
archivePrefix = {arXiv},
       eprint = {1201.1915},
 primaryClass = {astro-ph.SR},
       adsurl = {https://ui.adsabs.harvard.edu/abs/2012MNRAS.421.2820O}
}

@ARTICLE{Remillard2006,
       author = {{Remillard}, Ronald A. and {McClintock}, Jeffrey E.},
        title = "{X-Ray Properties of Black-Hole Binaries}",
      journal = {\araa},
         year = 2006,
        month = sep,
       volume = {44},
       number = {1},
        pages = {49-92},
          doi = {10.1146/annurev.astro.44.051905.092532},
archivePrefix = {arXiv},
       eprint = {astro-ph/0606352},
 primaryClass = {astro-ph},
       adsurl = {https://ui.adsabs.harvard.edu/abs/2006ARA&A..44...49R}
}

@ARTICLE{Brocksopp1999b,
       author = {{Brocksopp}, C. and {Fender}, R.~P. and {Larionov}, V. and {Lyuty}, V.~M. and {Tarasov}, A.~E. and {Pooley}, G.~G. and {Paciesas}, W.~S. and {Roche}, P.},
        title = "{Orbital, precessional and flaring variability of Cygnus X-1}",
      journal = {\mnras},
         year = {1999b},
        month = nov,
       volume = {309},
       number = {4},
        pages = {1063-1073},
          doi = {10.1046/j.1365-8711.1999.02919.x},
archivePrefix = {arXiv},
       eprint = {astro-ph/9906365},
 primaryClass = {astro-ph},
       adsurl = {https://ui.adsabs.harvard.edu/abs/1999MNRAS.309.1063B}
}

@ARTICLE{Tashiro2025,
       author = {{Tashiro}, Makoto and {Kelley}, Richard and {Watanabe}, Shin and {Maejima}, Hironori and {Reichenthal}, Lillian and {Toda}, Kenichi and {Hartz}, Leslie and {Santovincenzo}, Andrea and {Matsushita}, Kyoko and {Yamaguchi}, Hiroya and {Petre}, Robert and {Williams}, Brian and {Guainazzi}, Matteo and {Costantini}, Elisa and {Takei}, Yoh and {Ishisaki}, Yoshitaka and {Fujimoto}, Ryuichi and {Henegar-Leon}, Joy and {Sneiderman}, Gary and {Tomida}, Hiroshi and {Mori}, Koji and {Nakajima}, Hiroshi and {Terada}, Yukikatsu and {Holland}, Matthew and {Loewenstein}, Michael and {Miller}, Eric and {Sawada}, Makoto and {Kallman}, Timothy and {Kaastra}, Jelle and {Done}, Chris and {Enoto}, Teruaki and {Bamba}, Aya and {Corrales}, Lia and {Ueda}, Yoshihiro and {Kara}, Erin and {Zhuravleva}, Irina and {Fujita}, Yutaka and {Arai}, Yoshitaka and {Audard}, Marc and {Awaki}, Hisamitsu and {Ballhausen}, Ralf and {Baluta}, Chris and {Bando}, Nobutaka and {Behar}, Ehud and {Bialas}, Thomas and {Boissay-Malaquin}, Rozenn and {Brenneman}, Laura and {Brown}, Gregory V. and {Chiao}, Meng and {Cumbee}, Renata and {de Vries}, Cor and {den Herder}, Jan-Willem and {D{\'\i}az Trigo}, Mar{\'\i}a and {DiPirro}, Michael and {Dotani}, Tadayasu and {Carrero}, Jacobo Ebrero and {Ebisawa}, Ken and {Eckart}, Megan and {Eckert}, Dominique and {Eguchi}, Satoshi and {Ezoe}, Yuichiro and {Ferrigno}, Carlo and {Foster}, Adam and {Fukazawa}, Yasushi and {Fukushima}, Kotaro and {Furuzawa}, Akihiro and {Gallo}, Luigi C. and {Garcia Martinez}, Javier and {Gorter}, Nathalie and {Grim}, Martin and {Gu}, Liyi and {Hagino}, Kouichi and {Hamaguchi}, Kenji and {Hatsukade}, Isamu and {Hayashi}, Katsuhiro and {Hayashi}, Takayuki and {Hell}, Natalie and {Hodges-Kluck}, Edmund and {Horiuchi}, Takafumi and {Hornschemeier}, Ann and {Hoshino}, Akio and {Ichinohe}, Yuto and {Ikuta}, Chisato and {Iizuka}, Ryo and {Ishi}, Daiki and {Ishida}, Manabu and {Ishihama}, Naoki and {Ishikawa}, Kumi and {Ishimura}, Kosei and {Jaffe}, Tess and {Katsuda}, Satoru and {Kanemaru}, Yoshiaki and {Kenyon}, Steven and {Kilbourne}, Caroline and {Kimball}, Mark and {Kitamoto}, Shunji and {Kobayashi}, Shogo and {Kohmura}, Takayoshi and {Kubota}, Aya and {Leutenegger}, Maurice A. and {Maeda}, Yoshitomo and {Markevitch}, Maxim and {Matsumoto}, Hironori and {Matsuzaki}, Keiichi and {McCammon}, Dan and {McLaughlin}, Brian and {McNamara}, Brian and {Mernier}, Fran{\c{c}}ois and {Miko}, Joseph and {Miller}, Jon M. and {Minesugi}, Kenji and {Mitani}, Shinji and {Mitsuishi}, Ikuyuki and {Mizumoto}, Misaki and {Mizuno}, Tsunefumi and {Mukai}, Koji and {Murakami}, Hiroshi and {Mushotzky}, Richard and {Nakazawa}, Kazuhiro and {Natsukari}, Chikara and {Ness}, Jan-Uwe and {Nigo}, Kenichiro and {Nishiyama}, Mari and {Nobukawa}, Kumiko and {Nobukawa}, Masayoshi and {Noda}, Hirofumi and {Odaka}, Hirokazu and {Ogawa}, Mina and {Ogawa}, Shoji and {Ogorzalek}, Anna and {Okajima}, Takashi and {Okamoto}, Atsushi and {Ota}, Naomi and {Ozaki}, Masanobu and {Paltani}, Stephane and {Plucinsky}, Paul and {Porter}, F. Scott and {Pottschmidt}, Katja and {Quero}, Jose Antonio and {Sasaki}, Takahiro and {Sato}, Kosuke and {Sato}, Rie and {Sato}, Toshiki and {Sato}, Yoichi and {Seta}, Hiromi and {Shida}, Maki and {Shidatsu}, Megumi and {Shigeto}, Shuhei and {Shipman}, Russel and {Shinozaki}, Keisuke and {Shirron}, Peter and {Simionescu}, Aurora and {Smith}, Randall K. and {Soong}, Yang and {Suzuki}, Hiromasa and {Szymkowiak}, Andrew and {Takahashi}, Hiromitsu and {Takeo}, Mai and {Tamagawa}, Toru and {Tamura}, Keisuke and {Tanaka}, Takaaki and {Tanimoto}, Atsushi and {Terashima}, Yuichi and {Tsuboi}, Yohko and {Tsujimoto}, Masahiro and {Tsunemi}, Hiroshi and {Tsuru}, Takeshi Go and {Uchida}, Hiroyuki and {Uchida}, Nagomi and {Uchida}, Yuusuke and {Uchiyama}, Hideki and {Uno}, Shinichiro and {Vink}, Jacco and {Witthoeft}, Michael and {Wolfs}, Rob and {Yamada}, Satoshi and {Yamada}, Shinya and {Yamaoka}, Kazutaka and {Yamasaki}, Noriko and {Yamauchi}, Makoto and {Yamauchi}, Shigeo and {Yanagase}, Keiichi and {Yaqoob}, Tahir and {Yasuda}, Susumu and {Yoneyama}, Tomokage and {Yoshida}, Tessei and {Yukita}, Mihoko},
        title = "{X-Ray Imaging and Spectroscopy Mission}",
      journal = {\pasj},
         year = 2025,
        month = sep,
       volume = {77},
        pages = {S1-S9},
          doi = {10.1093/pasj/psaf023},
       adsurl = {https://ui.adsabs.harvard.edu/abs/2025PASJ...77S...1T}
}

@ARTICLE{Noda2025,
       author = {{Noda}, Hirofumi and {Mori}, Koji and {Tomida}, Hiroshi and {Nakajima}, Hiroshi and {Tanaka}, Takaaki and {Murakami}, Hiroshi and {Uchida}, Hiroyuki and {Suzuki}, Hiromasa and {Kobayashi}, Shogo Benjamin and {Yoneyama}, Tomokage and {Hagino}, Kouichi and {Nobukawa}, Kumiko and {Uchiyama}, Hideki and {Nobukawa}, Masayoshi and {Matsumoto}, Hironori and {Tsuru}, Takeshi Go and {Yamauchi}, Makoto and {Hatsukade}, Isamu and {Odaka}, Hirokazu and {Kohmura}, Takayoshi and {Yamaoka}, Kazutaka and {Yoshida}, Tessei and {Kanemaru}, Yoshiaki and {Hiraga}, Junko and {Dotani}, Tadayasu and {Ozaki}, Masanobu and {Tsunemi}, Hiroshi and {Sato}, Jin and {Takaki}, Toshiyuki and {Terada}, Yuta and {Miyazaki}, Keitaro and {Kusunoki}, Kohei and {Otsuka}, Yoshinori and {Yokosu}, Haruhiko and {Yonemaru}, Wakana and {Ichikawa}, Kazuhiro and {Nakano}, Hanako and {Takemoto}, Reo and {Matsushima}, Tsukasa and {Urase}, Reika and {Kurashima}, Jun and {Fuchi}, Kotomi and {Hayakawa}, Kaito and {Fukuda}, Masahiro and {Kamei}, Takamitsu and {Asahina}, Yoh and {Inoue}, Shun and {Amano}, Yuki and {Aoki}, Yuma and {Ito}, Yamato and {Kamatani}, Tomoya and {Takayama}, Kouta and {Sako}, Takashi and {Yoshimoto}, Marina and {Shima}, Kohei and {Higuchi}, Mayu and {Ninoyu}, Kaito and {Aoki}, Daiki and {Tsunomachi}, Shun and {Hayashida}, Kiyoshi},
        title = "{Soft X-ray Imager of the Xtend system on board XRISM}",
      journal = {\pasj},
         year = 2025,
        month = sep,
       volume = {77},
        pages = {S10-S22},
          doi = {10.1093/pasj/psaf011},
archivePrefix = {arXiv},
       eprint = {2502.08030},
 primaryClass = {astro-ph.IM},
       adsurl = {https://ui.adsabs.harvard.edu/abs/2025PASJ...77S..10N}
}

@ARTICLE{Xiang2011,
       author = {{Xiang}, Jingen and {Lee}, Julia C. and {Nowak}, Michael A. and {Wilms}, J{\"o}rn},
        title = "{Using the X-Ray Dust Scattering Halo of Cygnus X-1 to Determine Distance and Dust Distributions}",
      journal = {\apj},
         year = 2011,
        month = sep,
       volume = {738},
       number = {1},
          eid = {78},
        pages = {78},
          doi = {10.1088/0004-637X/738/1/78},
archivePrefix = {arXiv},
       eprint = {1106.3378},
 primaryClass = {astro-ph.HE},
       adsurl = {https://ui.adsabs.harvard.edu/abs/2011ApJ...738...78X}
}

@inproceedings{Ishisaki2022-ci,
 author = {Ishisaki, Yoshitaka and Kelley, Richard L and Awaki, Hisamitsu
and Balleza, Jesus C and Barnstable, Kim R and Bialas, Thomas G
and Boissay-Malaquin, Rozenn and Brown, Gregory V and Canavan,
Edgar R and Cumbee, Renata S and Carnahan, Timothy M and Chiao,
Meng P and Comber, Brian J and Costantini, Elisa and den Herder,
Jan-Willem A and Dercksen, Johannes and de Vries, Cor P and
DiPirro, Michael J and Eckart, Megan E and Ezoe, Yuichiro and
Ferrigno, Carlo and Fujimoto, Ryuichi and Gorter, Nathalie and
Graham, Steven M and Grim, Martin and Hartz, Leslie S and
Hayakawa, Ryota and Hayashi, Takayuki and Hell, Natalie and
Hoshino, Akio and Ichinohe, Yuto and Ishida, Manabu and Ishikawa,
Kumi and James, Bryan L and Kenyon, Steven J and Kilbourne,
Caroline A and Kimball, Mark O and Kitamoto, Shunji and
Leutenegger, Maurice A and Maeda, Yoshitomo and McCammon, Dan and
Miko, Joseph J and Mizumoto, Misaki and Okajima, Takashi and
Okamoto, Atsushi and Paltani, Stephane and Porter, Frederick S
and Sato, Kosuke and Sato, Toshiki and Sawada, Makoto and
Shinozaki, Keisuke and Shipman, Russell and Shirron, Peter J and
Sneiderman, Gary A and Soong, Yang and Szymkiewicz, Richard and
Szymkowiak, Andrew E and Takei, Yoh and Tamura, Keisuke and
Tsujimoto, Masahiro and Uchida, Yuusuke and Wasserzug, Stephen
and Witthoeft, Michael C and Wolfs, Rob and Yamada, Shinya and
Yasuda, Susumu},
 booktitle = {Space Telescopes and Instrumentation 2022: Ultraviolet to Gamma
Ray},
 doi = {10.1117/12.2630654},
 editor = {den Herder, Jan-Willem A and Nakazawa, Kazuhiro and Nikzad,
Shouleh},
 month = {31~August},
 pages = {121811S},
 publisher = {SPIE},
 series = {Society of Photo-Optical Instrumentation Engineers (SPIE)
Conference Series},
 title = {Status of resolve instrument onboard x-ray imaging and
spectroscopy mission ({XRISM})},
 url = {https://www.spiedigitallibrary.org/conference-proceedings-of-spie/12181/2630654/Status-of-resolve-instrument-onboard-x-ray-imaging-and-spectroscopy/10.1117/12.2630654.full},
 volume = {12181},
 year = {2022}
}

@inproceedings{Midooka2020-ns,
 author = {Midooka, Takuya and Tsujimoto, Masahiro and Kitamoto, Shunji and
Nakaniwa, Nozomi and Maeda, Yoshitomo and Hayakawa, Shinjiro and
Ishida, Manabu and Ebisawa, Ken and Tominaga, Mayu},
 booktitle = {Space Telescopes and Instrumentation 2020: Ultraviolet to Gamma
Ray},
 doi = {10.1117/12.2559451},
 editor = {den Herder, Jan-Willem A and Nakazawa, Kazuhiro and Nikzad,
Shouleh},
 isbn = {9781510636750,9781510636767},
 month = {13~December},
 publisher = {SPIE},
 title = {{X}-ray transmission measurements of the gate valve for the x-ray
astronomy satellite {XRISM}},
 url = {https://www.spiedigitallibrary.org/conference-proceedings-of-spie/11444/114445C/X-ray-transmission-measurements-of-the-gate-valve-for-the/10.1117/12.2559451.short},
 year = {2020}
}

@inproceedings{Mori2022-wo,
 author = {Mori, Koji and Tomida, Hiroshi and Nakajima, Hiroshi and Okajima,
Takashi and Noda, Hirofumi and Tanaka, Takaaki and Uchida,
Hiroyuki and Hagino, Kouichi and Kobayashi, Shogo Benjamin and
Suzuki, Hiromasa and Yoshida, Tessei and Murakami, Hiroshi and
Uchiyama, Hideki and Nobukawa, Masayoshi and Nobukawa, Kumiko and
Yoneyama, Tomokage and Matsumoto, Hironori and Tsuru, Takeshi and
Yamauchi, Makoto and Hatsukade, Isamu and Ishida, Manabu and
Maeda, Yoshitomo and Hayashi, Takayuki and Tamura, Keisuke and
Boissay-Malaquin, Rozenn and Sato, Toshiki and Hiraga, Junko and
Kohmura, Takayoshi and Yamaoka, Kazutaka and Dotani, Tadayasu and
Ozaki, Masanobu and Tsunemi, Hiroshi and Kanemaru, Yoshiaki and
Sato, Jin and Takaki, Toshiyuki and Terada, Yuta and Miyazaki,
Keitaro and Kusunoki, Kohei and Otsuka, Yoshinori and Yokosu,
Haruhiko and Yonemaru, Wakana and Asahina, Yoh and Asakura,
Kazunori and Yoshimoto, Marina and Ode, Yuichi and Sato, Junya
and Hakamata, Tomohiro and Aoyagi, Mio and Aoki, Yuma and
Tsunomachi, Shun and Doi, Toshiki and Aoki, Daiki and Fujisawa,
Kaito and Kitajima, Masatoshi and Hayashida, Kiyoshi},
 booktitle = {Space Telescopes and Instrumentation 2022: Ultraviolet to Gamma
Ray},
 doi = {10.1117/12.2626894},
 editor = {den Herder, Jan-Willem A and Nikzad, Shouleh and Nakazawa,
Kazuhiro},
 month = {August},
 pages = {121811T},
 publisher = {Proceedings of SPIE Astronomical Telescopes and Instrumentation
2022},
 series = {Society of Photo-Optical Instrumentation Engineers (SPIE)
Conference Series},
 title = {Xtend, the soft x-ray imaging telescope for the {X}-Ray Imaging
and Spectroscopy Mission ({XRISM})},
 url = {http://dx.doi.org/10.1117/12.2626894},
 volume = {12181},
 year = {2022}
}

@incollection{Sato2023-gr,
 address = {Singapore},
 author = {Sato, Kosuke and Uchida, Yuusuke and Ishikawa, Kumi},
 booktitle = {High-Resolution X-ray Spectroscopy},
 doi = {10.1007/978-981-99-4409-5\_5},
 isbn = {9789819944088,9789819944095},
 issn = {2731-734X,2731-7358},
 pages = {93--123},
 publisher = {Springer Nature Singapore},
 title = {Hitomi/{XRISM} Micro-Calorimeter},
 url = {https://link.springer.com/10.1007/978-981-99-4409-5_5},
 year = {2023}
}

@inproceedings{Tashiro2020-iz,
 author = {Tashiro, Makoto S and Maejima, Hironori and Toda, Kenichi and
Kelley, Richard L and Reichenthal, Lillian and Hartz, Leslie and
Petre, Robert and Williams, Brian J and Guainazzi, Matteo and
Costantini, Elisa and Fujimoto, Ryuichi and Hayashida, Kiyoshi
and Henegar-Leon, Joy and Holland, Matt and Ishisaki, Yoshitaka
and Kilbourne, Caroline and Loewenstein, Mike and Matsushita,
Kyoko and Mori, Koji and Okajima, Takashi and Porter, F Scott and
Sneiderman, Gary and Takei, Yoh and Terada, Yukikatsu and Tomida,
Hiroshi and Yamaguchi, Hiroya and Watanabe, Shin and Akamatsu,
Hiroki and Arai, Yoshitaka and Audard, Marc and Awaki, Hisamitsu
and Babyk, Iurii and Bamba, Aya and Bando, Nobutaka and Behar,
Ehud and Bialas, Thomas and Boissay-Malaquin, Rozenn and
Brenneman, Laura and Brown, Greg and Canavan, Edgar and Chiao,
Meng and Comber, Brian and Corrales, Lia and Cumbee, Renata and
de Vries, Cor and den Herder, Jan-Willem and Dercksen, Johannes
and Diaz-Trigo, Maria and DiPirro, Michael and Done, Chris and
Dotani, Tadayasu and Ebisawa, Ken and Eckart, Megan and Eckert,
Dominique and Eguchi, Satoshi and Enoto, Teruaki and Ezoe,
Yuichiro and Ferrigno, Carlo and Fujita, Yutaka and Fukazawa,
Yasushi and Furuzawa, Akihiro and Gallo, Luigi and Gorter,
Nathalie and Grim, Martin and Gu, Liyi and Hagino, Kouichi and
Hamaguchi, Kenji and Hatsukade, Isamu and Hawthorn, David and
Hayashi, Katsuhiro and Hell, Natalie and Hiraga, Junko and
Hodges-Kluck, Edmund and Horiuchi, Takafumi and Hornschemeier,
Ann and Hoshino, Akio and Ichinohe, Yuto and Iga, Sayuri and
Iizuka, Ryo and Ishida, Manabu and Ishihama, Naoki and Ishikawa,
Kumi and Ishimura, Kosei and Jaffe, Tess and Kaastra, Jelle and
Kallman, Timothy and Kara, Erin and Katsuda, Satoru and Kenyon,
Steven and Kimball, Mark and Kitaguti, Takao and Kitamoto, Shunji
and Kobayashi, Shogo and Kobayashi, Akihide and Kohmura,
Takayoshi and Kubota, Aya and Leutenegger, Maurice and Li, Muzi
and Lockard, Tom and Maeda, Yoshitomo and Markevitch, Maxim and
Martz, Connor and Matsumoto, Hironori and Matsuzaki, Keiichi and
McCammon, Dan and McLaughlin, Brian and McNamara, Brian and Miko,
Joseph and Miller, Eric and Miller, Jon and Minesugi, Kenji and
Mitani, Shinji and Mitsuishi, Ikuyuki and Mizumoto, Misaki and
Mizuno, Tsunefumi and Mukai, Koji and Murakami, Hiroshi and
Mushotzky, Richard and Nakajima, Hiroshi and Nakamura, Hideto and
Nakazawa, Kazuhiro and Natsukari, Chikara and Nigo, Kenichiro and
Nishioka, Yusuke and Nobukawa, Kumiko and Nobukawa, Masayoshi and
Noda, Hirofumi and Odaka, Hirokazu and Ogawa, Mina and Ohashi,
Takaya and Ohno, Masahiro and Ohta, Masayuki and Okamoto, Atsushi
and Ota, Naomi and Ozaki, Masanobu and Paltani, Stephane and
Plucinsky, Paul and Pottschmidt, Katja and Sampson, Michael and
Sasaki, Takahiro and Sato, Kosuke and Sato, Rie and Sato, Toshiki
and Sawada, Makoto and Seta, Hiromi and Shibano, Yasuko and
Shida, Maki and Shidatsu, Megumi and Shigeto, Shuhei and
Shinozaki, Keisuke and Shirron, Peter and Simionescu, Aurora and
Smith, Randall and Someya, Kazunori and Soong, Yang and Sugawara,
Keisuke and Sugawara, Yasuharu and Szymkowiak, Andy and
Takahashi, Hiromitsu and Takeshima, Toshiaki and Tamagawa, Toru
and Tamura, Keisuke and Tanaka, Takaaki and Tanimoto, Atsushi and
Terashima, Yuichi and Tsuboi, Yohko and Tsujimoto, Masahiro and
Tsunemi, Hiroshi and Tsuru, Takeshi and Uchida, Hiroyuki and
Uchida, Yuusuke and Uchiyama, Hideki and Ueda, Yoshihiro and Uno,
Shinichiro and Vink, Jacco and Watanabe, Tomomi and Wittheof,
Michael and Wolfs, Rob and Yamada, Shinya and Yamaoka, Kazutaka
and Yamasaki, Noriko and Yamauchi, Makoto and Yamauchi, Shigeo
and Yanagase, Keiichi and Yaqoob, Tahir and Yasuda, Susumu and
Yoshida, Tessei and Yoshioka, Nasa and Zhuravleva, Irina},
 booktitle = {Space Telescopes and Instrumentation 2020: Ultraviolet to Gamma
Ray},
 doi = {10.1117/12.2565812},
 editor = {den Herder, Jan-Willem A and Nakazawa, Kazuhiro and Nikzad,
Shouleh},
 isbn = {9781510636750,9781510636767},
 month = {13~December},
 pages = {1144422},
 publisher = {SPIE},
 series = {Society of Photo-Optical Instrumentation Engineers (SPIE)
Conference Series},
 title = {Status of x-ray imaging and spectroscopy mission ({XRISM})},
 url = {https://www.spiedigitallibrary.org/conference-proceedings-of-spie/11444/2565812/Status-of-x-ray-imaging-and-spectroscopy-mission-XRISM/10.1117/12.2565812.full},
 volume = {11444},
 year = {2020}
}

\end{document}